\documentclass{article}

\usepackage[letterpaper,top=2cm,bottom=2cm,left=2cm,right=2cm,marginparwidth=1.75cm]{geometry}
\usepackage{amsmath, bm}
\usepackage{amssymb}
\usepackage[bb=dsserif]{mathalpha} 
\usepackage[colorlinks=true, allcolors=blue]{hyperref}
\usepackage{graphicx} 
\usepackage{xcolor}
\usepackage{longtable}
\usepackage{booktabs}
\usepackage{array}
\usepackage{authblk}
\usepackage{pdflscape}
\usepackage{afterpage}
\usepackage[skip=10pt plus1pt]{parskip}
\usepackage{setspace}
\newcommand{\code}[1]{\texttt{#1}}
\usepackage{capt-of}
\usepackage[backend=biber,autocite=superscript]{biblatex}
\let\parencite\supercite
\title{Collaborative estimation and evaluation of SARS-CoV-2 variant nowcasting in the United States}

\author[1,$\dagger$]{Isaac MacArthur}
\author[1,$\dagger$]{Thomas Robacker}
\author[2]{Bren Case}
\author[3]{Spencer J. Fox}
\author[4]{Dylan H. Morris}
\author[1] {Evan L. Ray}
\author[1]{Benjamin Rogers}
\author[1]{Becky Sweger}
\author[5]{Natalie M. Linton}
\author[6]{John Huddleston}
\author[4]{Andrew Magee}
\author[4]{Zachary Susswein}
\author[6]{Jover Lee}
\author[6,7]{Trevor Bedford}
\author[6]{Marlin D. Figgins}
\author[8]{Ehsan Suez}
\author[3]{Rajath Prabhakar}
\author[5]{Tom{\'a}s M. Le{\'o}n}
\author[5]{Brent Siegel}
\author[5]{Mugdha Thakur}
\author[5]{Christopher M. Hoover}
\author[5]{Rahil Ryder}
\author[5]{Jesse Elder}
\author[9]{Michael Kupperman}
\author[9]{Ruian Ke}
\author[9]{Emma Goldberg}
\author[10]{Sebastian Funk}
\author[1]{Maryclare Griffin}
\author[11]{Nicholas G. Reich}
\author[10, *]{Kaitlyn E. Johnson}

\affil[1]{Department of Mathematics and Statistics, University of Massachusetts, Amherst, United States of America}
\affil[2]{Epidemiology \& Biostatistics, University of Georgia, Athens, GA}
\affil[3]{School of Informatics, Computing, and Cybersystems, Northern Arizona University, Flagstaff, Arizona}
\affil[4]{The Center for Forecasting and Outbreak Analytics, Centers for Disease Control and Prevention}
\affil[5]{California Department of Public Health}
\affil[6]{Vaccine and Infectious Disease Division, Fred Hutchinson Cancer Center, Seattle, Washington}
\affil[7]{Howard Hughes Medical Institute, Seattle, Washington}
\affil[8]{Institute of Bioinformatics, University of Georgia, Athens, Georgia}
\affil[9]{Theoretical Biology and Biophysics, Los Alamos National Laboratory, New Mexico}
\affil[10]{Department of Infectious Disease Epidemiology \& Dynamics, London School of Hygiene \& Tropical Medicine}
\affil[11]{School of Public Health and Health Sciences, University of Massachusetts, Amherst, United States of America}

\affil[$\dagger$]{These authors contributed equally to this work.}
\affil[*] {Corresponding author: kaitlyn.johnson@lshtm.ac.uk}

\begin{document}
\maketitle
\begin{abstract}
 The ability to estimate and predict pathogen variant dynamics can inform public health responses, including planning for increased transmission or severity, shifts in population immunity, or changes to vaccine or therapeutic effectiveness. The COVID-19 pandemic demonstrated the importance of monitoring SARS-CoV-2 variant evolution through viral genome sequencing, enabling predictive models to estimate variant frequencies in the recent past, present, and short-term future. Collaborative forecasting Hubs provided a valuable way to centralize predictive modeling of epidemiological indicators such as cases, hospitalizations, and deaths during the pandemic; however, none existed for variant dynamics. Here, we discuss the creation of the United States SARS-CoV-2 Variant Nowcast Hub, designed to solicit estimates of the relative abundance of a specified set of SARS-CoV-2 variants at the U.S. state level. We discuss the design decisions and challenges in building the Hub and its scoring procedures. Using submissions from the Hub's first respiratory virus season (nowcast dates October 9th, 2024 to June 4th, 2025), we evaluate five individual models and a baseline model. We found that the baseline model, which pools sequences across the U.S., performs well overall, with most individual models performing similarly or slightly worse. Locations with lower sequencing volumes exhibited greater variability in model performance. Models submitted for a single location outperformed those submitted for all locations, potentially due to greater timeliness and magnitude of local data. Much remains to be investigated regarding relative model performance across different phases of variant emergence, and we conclude by proposing future directions within and beyond this Hub.

\end{abstract}

\section*{Introduction}
The COVID-19 pandemic led to a massive increase in the use and availability of genomic sequencing for viral pathogens. Viral pathogen genomic data can provide a number of insights including understanding the origins of a virus \parencite{Crits-Christoph2024Market, Hill2022Phylogenetics}, reconstructing transmission events \parencite{Campbell2019Bayesian, Giardina2022Methods}, and understanding and predicting pathogen evolutionary and transmission dynamics \parencite{Molan2025}. The emergence of SARS-CoV-2 strains with transmission and immune evasion advantages, such as Delta and Omicron, demonstrated the value of being able to estimate and forecast pathogen variant dynamics in real time \parencite{Volz2021, daviesesttransmissibility2021, Taylor2025, Figgins2025, Park2023, tegally2022emergence, Figgins2024, Feng2024, Reichmuth2023}, as these informed predictions of variant-driven surges of transmission \parencite{Johnson2021} and could be used to inform policy regarding antivirals \parencite{Ragonnet-Cronin2023} and vaccine strain selection \parencite{Huddleston2020Forecasting, Lee2025Reproducible, Krammer2023COVID}. The utility of modeling variant dynamics extends beyond COVID-19, having important implications for influenza \parencite{hay2025evaluation}, dengue \parencite{Lorenz2025}, mpox \parencite{Srivastava2025}, Ebola \parencite{McCrone2025}, and norovirus \parencite{Barclay2025}. Recent efforts have increased the availability of open-source viral pathogen genomic data in pathogens beyond COVID-19 \parencite{benson2012genbank} and influenza \parencite{pathoplexus2024}. Reliable nowcasts and short-term forecasts of variant dynamics may improve the utility of viral genomic data for decision making. 

This work is based on the U.S. SARS-CoV-2 Variant Nowcast Hub \parencite{Variant-nowcast2024} (hereinafter referred to as 'the Hub'), a collaborative platform for real-time estimation of the relative prevalence of variants (a specified set of Nextstrain \parencite{Hadfield2018, andrews2026nextstrain} clades) across time and U.S. jurisdictions. In this context, a "nowcast" refers to all of the following: a "hindcast", where predictions are made for the past, a true "nowcast" of an estimate of the current time, and a 10-day ahead short-term forecast.

Producing reliable estimates of variant frequency dynamics can be challenging for a number of reasons. While the availability of genomic sequencing data has increased \parencite{Chen2022}, it is still much less abundant than traditional epidemiological indicators such as counts of incident cases or hospital admissions, as typically a small proportion of positive cases are chosen for subsequent genomic sequencing. Processing delays from specimen collection to publication of sequence data mean that data is even more sparse for the most recent time points. This can obscure recent trends, make interpreting the data difficult, and pose challenges for simple models due to a lack of recent data and the increased likelihood of more extreme-valued proportions due to small counts. 

In addition to the methodological challenge of producing accurate variant nowcasts/forecasts, evaluating the performance of these models retrospectively is  difficult for a number of reasons.  First, the quantity of epidemiological interest, relative variant proportions, is observed through sample proportions, whose variability depends on the final number of specimens collected for sequencing per day. This number is unknown at the time nowcasts/forecasts of variant proportions are submitted due to lags from specimen collection to sequencing. A second challenge is the dynamic nature of the prediction target, the variants of interest, which must be specified each week and change as the pathogen evolves, requiring both Hub administrators to specify the variant targets and modelers to adapt their output structure to reflect these targets, each week. This is in contrast to other forecast hub prediction targets such as test-positive cases or test-positive hospital admissions, which, while they may change as the reporting landscape shifts, do not change on a week-to-week basis. 
Third, when observations of proportions are partially observed at the time of nowcast submissions, it could be to the modeler's advantage to predict a value that is close to the frequency observed at the time the forecast is made, thus deviating from the underlying (likely smooth) function the model would predict in the absence of data. This requires evaluation procedures that address this incentive and ensure forecasts of the true predictive distribution are incentivized. Fourth, evaluation is made complicated by the fact that the clade assignment model changes over time \parencite{Aksamentov2021}. As more sequences become available and as SARS-CoV-2 continues to evolve, the way that branches, defined by key mutational signatures, are grouped to form clades (a group of sequences defined by a specific set of mutational signatures), changes in time. This phylogenetic nomenclature system is what allows us to define the signature for a new lineage via a new clade designation. However, it also means that a sequence that at one time was assigned to a particular clade may at a later time be assigned to a different clade at a later time. This usually occurs as existing clades are split to designate derived daughter clades, e.g. in April 2025 clade 25B (pango lineage NB.1.8.1) was designated as daughter to clade 24D (pango lineage XDV.1) and viruses previously classified as 24D became a mix of 24D and 25B viruses. This differs from other nowcasting or forecasting hubs where teams are predicting expected observations that can be directly compared to the observations.

Despite these challenges, the recognized utility of variant dynamics estimates during the COVID-19 pandemic resulted in the creation of a number of dashboards \parencite{Paul2021, CDCVariantNowcast, COGUKConsortium, nextstrainforecasts}, reports \parencite{UKHSATechnicalBriefing38}, and public-facing tools for variant nowcasting \parencite{CoVSpectrum}. For the most part, the variant dynamics modeling components of these public-facing outputs were estimated from a single model chosen by the developers, making it difficult to identify and evaluate which types of models perform best under different circumstances. 

Collaborative forecasting and nowcasting can improve both model evaluability and integration of modeling outputs into public health practice by centralizing and coordinating predictive modeling efforts. Standard targets facilitate model evaluation and enable assessment of the reliability of model estimates \parencite{hubverse2025, reich2022collaborative, cramer2022evaluation}. For nowcasting and forecasting of traditional epidemiological indicators such as cases, hospital admissions, and deaths, there are a number of ``templates'' for how to build a collaborative Hub that have been used across a wide range of pathogens and geographic settings \parencite{cramer2022evaluation, lopez2024challenges, Reich2019Influenza, reich2022collaborative, Ray2023, Bracher2021, hubverse2025}  with standardized tools having been developed to ensure compatibility both within and across Hubs \parencite{hubverse2025, hubverse2025github}. However, all of these Hubs solicit predictions of pre-specified quantities such as the number of hospital admissions, cases, and deaths, or the percent of all cases due to a particular pathogen or syndrome. In this project, we solicited estimates of a changing set of targets corresponding to the currently circulating variants, which modelers have to update each week. This new forecasting target required that we address a number of open questions regarding how to standardize targets, what output format to solicit from modelers, and what process and metrics to use for evaluating these outputs. While a number of efforts have been made to evaluate point predictions of variant proportions using the Brier score \parencite{Susswein2023} and mean absolute error \parencite{abousamra2024fitness}, we are aiming to evaluate the full probabilistic distribution of variant proportions. Additionally, no previous work has evaluated model performance with real-time outputs in which clade assignment models shift over time. While the focus of this work is not on justifying the choice of evaluation metric and procedure, this work is the first to our knowledge to propose a path forward for standardized real-time continuous evaluation of estimated latent clade proportions. 

In this paper, we describe the evaluation framework used in the Hub and apply it to real-time outputs submitted to the Hub. 
The paper is structured as follows: 
Section 1 of the methods describes the data that is generated and stored by the Hub and the data submitted by modeling teams, 
Section 2 of the methods provides an overview of the models submitted to the Hub during this period of initial assessment, from October 9th, 2024 to June 4th, 2025, and Section 3 of the methods describes how model performance is evaluated. 
In the results section, we start by providing an overview of the final genomic sequencing data available from the assessment period. We compare the data available to modelers when nowcasts were solicited to what was available at the time of evaluation. Finally, we summarize model performance---overall, by horizon, by location, and by nowcast date. We also examine model performance in a few jurisdictions during a period of variant emergence: the displacement of clade 24F by clade 25A. A real-time interactive dashboard showing nowcasts compared to later observed data has been developed to facilitate the communication of these outputs to relevant stakeholders in real-time and retrospectively (\url{https://reichlab.io/variant-nowcast-hub-dashboard/explore.html}). A key reason for building and launching this Hub is to explore useful ways to evaluate and communicate variant dynamics in a collaborative modeling effort, with the potential to expand beyond SARS-CoV-2 to other rapidly evolving pathogens, and we conclude by proposing future work related to SARS-CoV-2 pathogen evolution and beyond.

\section{Methods}

\subsection{Data created and stored by the Hub}
\label{sec:datacreatedandstored}
The Hub creates and collects data to facilitate the smooth running of modeling rounds; below are descriptions of a few of the data files that are stored or created by the Hub.
\subsubsection{Clade list}
\label{sec:cladelist}
 
 In contrast to other nowcasting and forecasting Hubs, the prediction targets for a variant nowcasting Hub are dynamically defined, meaning that the Hub itself has to specify which variants to solicit estimates for. Nextstrain clades \parencite{Hadfield2018, andrews2026nextstrain} were chosen to define target variants because they reflect large-scale trends in SARS-CoV-2 evolution \parencite{Telenti2022} with 58 clades defined from the beginning of the pandemic to Feb 2026. A suitably-sized set of relevant clades to model can be obtained with a simple frequency cutoff. Pango lineages represent another, more fine-grained approach to naming SARS-CoV-2 variants \parencite{rambaut2020dynamic} with $>$4700 lineages designated as of Feb 2026. Selecting a suitable number of relevant variants for nowcasting and forecasting from Pango lineages is not straightforward, as overly coarse variant designation will miss biologically relevant differences while overly granular designation will result in overly sparse data with too few observations per variant. Both systems are inherently phylogenetic and specify relationships between parent/child clades and parent/child lineages. Additionally, there is typically sufficient correspondence between them that it is possible to model Nextstrain clades but discuss results in terms of Pango lineages, as shown in Romer and Neher \parencite{Roemer2024Tree}. For example, Nextstrain clade 24A corresponds to Pango lineage JN.1, 22F to XBB, and 21L to BA.2.1. In order to facilitate interpretation and translation of outputs, we use a mapping from Nextstrain clades to Pango lineages provided by Nextstrain (\url{https://github.com/nextstrain/ncov/blob/master/defaults/clade_display_names.yml}) in our visualizations of the model predictions. We will use the term ``clade'' to describe the set of designated Nextstrain clades that the Hub solicits predictions for, and use the term ``variant'' to describe more generally groupings of pathogen sequences by their mutational signatures. 

Early Monday morning ($\sim$ 3am United States Eastern Standard Time) prior to a Wednesday on which submissions are due, the Hub generates a JSON file with two high-level properties:
    \begin{itemize}
        \item \code{clades}: an array of Nextclade clade names that will be accepted in submission files for the upcoming deadline.
        \item \code{meta}: metadata relevant to the upcoming round, including links to the Nextstrain sequence information, the Nextclade dataset used to generate the above clades array, and the total number of sequences from the last three weeks of data used to generate the round's clade list.  
    \end{itemize}
This clade selection is based on the ``full open'' Nextstrain \parencite{Hadfield2018, andrews2026nextstrain} sequence metadata files, which are curated versions of data sourced from the National Center for Biotechnology Information (NCBI) GenBank \parencite{benson2012genbank}, the COVID-19 Genomics U.K. Consortium (CoG-UK) and the Robert Koch Institute(RKI), and include labeling of Nextstrain clade. The Nextstrain files are updated once per week on Saturday. Throughout this work, we use ``the GenBank data'' to refer to the NCBI GenBank data curated by Nextstrain. The Hub pulls the most recent version of the file when the workflow runs each week on Monday morning. The precise clade assignment model that was used as well as the version of the raw sequence data are stored as metadata. The clade assignment model is used to assign a sequence to a particular clade. As time goes on, this ``reference tree" evolves to form new clades based on new ``branches'' or groups of branches of the tree. Thus, to facilitate reproducibility and evaluation, it is critical to track the version of the reference tree used for clade assignment prior to the date the nowcast was made. This way, the Hub can both reproduce the counts of sequences by clade available as of the nowcast date as well as the counts of sequences made available later, once the majority of the sequences collected before, during, and shortly after the nowcast date have been submitted to the GenBank database. 

To generate the set of clades for teams to model, the Hub selects up to nine individual clades with the highest observed frequency over the last three complete USA/CDC epidemiological weeks (a.k.a. MMWR weeks) \parencite{MMWR} preceding the Wednesday submission date that meet the following criteria:
\begin{itemize}
    \item Accounts for at least one percent of observations across the U.S. in any of the three complete USA/CDC epidemiological weeks.
    \item  (After April 16th, 20205): Has had at least two sequences present in the data across the last three complete USA/CDC epidemiological weeks.
\end{itemize}

Any clades which do not meet these criteria are grouped into an ``other'' category for which predictions of combined prevalence are also collected. No more than 10 clades (including ``other'') are selected in a given week.

This algorithm was designed to ensure that the resulting list of clades for modeling consistently produced a reasonable number of distinct clades as modeling targets while having a high likelihood of capturing all clades that were of "epidemiological interest" over the nowcasting period (i.e. we did not want to miss an emerging clade, but we also did not want to force teams to model many very low frequency clades of little importance). Experimentation showed this to be approximately 10 clades or fewer in order to accommodate a sufficient number of samples of the frequency of every clade in each location for six weeks. Another goal was to ensure clade selection was algorithmic and operated without human intervention. An algorithmic approach makes the choice of targets more transparent and simplifies Hub administration. 

\subsubsection{Target data}
\label{sec:targetdata}
 The Hub provides target data in the form of counts of the number of sequences of each designated clade in the GenBank data, indexed by specimen collection date. A version of this data set is created each week with data that was available as of the nowcast date, ranging from the 90 days before the nowcast date up until the nowcast date. In practice, data is rarely available for the week or two before the nowcast date. It is anticipated that most modelers will use extended historical data beyond 90 days in the past. Modelers are free to use any additional data sources they have access to. 
 
\subsection{Evaluation data: using the clade assignment as of the nowcast date for evaluation}
\label{sec:evaluationdata}

To generate an evaluation dataset for a particular nowcast date, the Hub extracts all of the sequences collected during the prediction horizon period for that nowcast date (31 days before and 10 days after the nowcast date). Sequences are extracted 90 days after the nowcast date. Historical data show that 90 days is sufficient time for nearly all sequences that have been collected during the prediction horizon to be reported (Fig. S1). Sequence assignments are made by assigning all the sequences collected during the prediction horizon, including those that were reported after the nowcast date, using the reference tree that was available as of the nowcast date. We created an open source python package, CladeTime \parencite{cladetime}, to perform these retrospective clade assignments. This means that each sequence is assigned the clade it would have been had it been reported by the nowcast date.

\begin{figure}[h]
    \centering
    \includegraphics[width=1\textwidth]{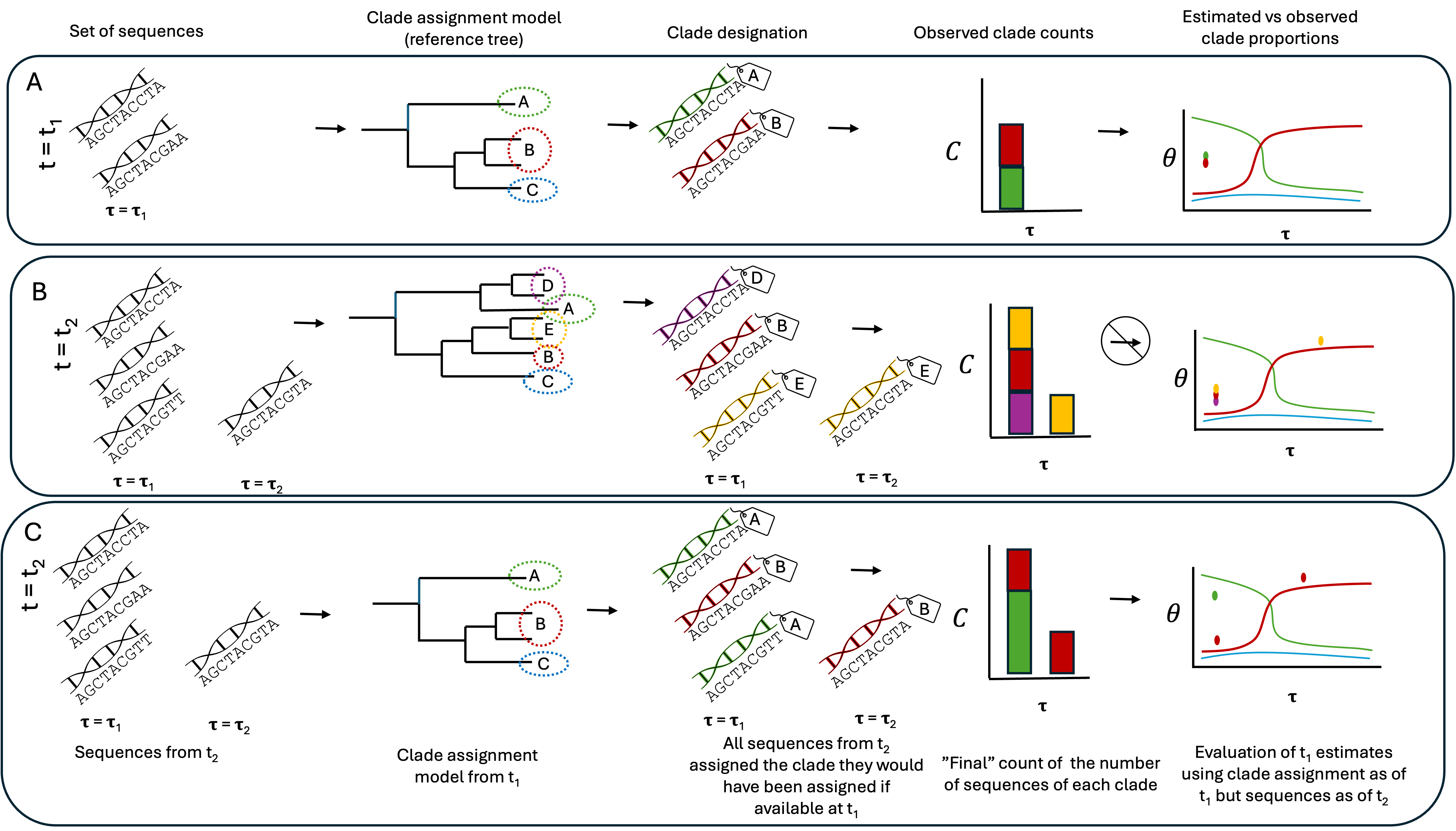}
    \caption{Illustration of clade assignment using different combinations of dates when sequences are available and versions of the reference tree. Two dates are relevant: the nowcast date, $t_1$, and the evaluation date, $t_2$, where as $\tau$ describes the specimen collection dates. (A). Clade assignment of sequences reported as of the nowcast date $t_1$, using the reference tree available as of the nowcast date; used for model estimation. (B). Clade assignment of sequences reported as of the evaluation date $t_2$ using the reference tree available as of the evaluation date; this is not used. (C). Clade assignment of sequences reported as of the evaluation date, $t_2$ using the reference tree available as of the nowcast date; used for nowcast evaluation. Clade assignments using different reference trees are not comparable, so the same reference tree is used when creating data for model fitting and for evaluation (A and C).
}
    \label{fig:schematiccladeassignment}
\end{figure}

In the formation of this Hub, we specifically chose not to elicit forecasts for new clades, because the emergence of and naming of new clades is highly stochastic and would likely require modeling the evolution of mutational signatures at the sequence level over larger time scales. We instead are focused on eliciting estimates of relative abundance of the currently circulating clades for recent and near-term future time periods. For this reason, our target for evaluation is to assess the performance of these estimates given the sequences available at a later time and the clade they would have been assigned had the sequences been available at the time of the nowcast, as is described in \autoref{fig:schematiccladeassignment}.

\subsection{What teams submit (model outputs)}
\label{sec:modeloutputs}
Teams are asked to predict the proportion of all infections of SARS-CoV-2 of each specified clade throughout the U.S., at a daily timescale and the geographic resolution of all 50 U.S. states plus Washington, D.C. and Puerto Rico. Teams are not required to submit for all locations. The Hub does not solicit estimates for the U.S. as a whole, since evaluating this quantity is not straightforward due to the heterogeneity in levels of infections and sequencing across jurisdictions. Details about these choices follow in subsections below.

\subsubsection{The forecast target: latent clade proportions}
\label{sec:forecasttarget}
Our choice in defining the forecast target of latent clade proportions relies on the assumption that the quantity that epidemiologists care about most is the relative prevalence (proportion) of all infections of each clade circulating in the population. This quantity is not directly observable; what we observe are noisy counts of the number of sequences designated to a particular clade based on a relatively small sample of sequenced cases. The variance of those counts, as well as the observed frequencies calculated based on them, depends on the final number of sequences collected. However, we assume that the final number of sequences collected is not of as much epidemiological interest as the latent variant proportions. In the context of the Hub, we therefore do not ask modelers to forecast the number of sequences collected at each date and location. Additionally, the Hub is focused on soliciting predictions of clade proportions among a pre-specified set of designated clades circulating at the time the nowcast is solicited (See Section \ref{sec:cladelist} for more details). Predictions of the emergence and/or designation of new clades are out of scope for this hub. 

The Hub therefore solicits predictions $\hat{\theta}_{l,t}$ for true latent clade proportions $\theta_{l,t}$ for a particular specimen collection date $t$ and location $l$. The predictions $\hat{\theta}_{l,t}$ are $K$-vectors of continuous numbers between 0 and 1 which are drawn from a probability simplex (i.e., the values sum to one, ${\theta_{k,l,t} \in \mathbb{R}^K: \theta_{k,l,t} \geq 0  \; \text{and} \sum_k \theta_{k,l,t} = 1}$ for all $l$ and $t$), where $K$ is the number of clades specified. The $k^{\text{th}}$ element $ \hat{\theta}_{k,l,t}$ is the estimated true proportion of all SARS-CoV-2 infections which are clade $k$ in a particular location on a particular date, corresponding to the specimen collection date. We refer to these as estimates of the latent clade proportions as they are not directly observed. We observe $C_{k, l,t} = (C_{1, l,t}, \ldots, C_{k, l, t})$, a vector of counts for each of the $K$ clades in a particular location on a particular specimen collection date, where $N_{l,t} = \sum_{k} C_{k,l,t}$ is the total number of sequences collected in that location on that target date.  However, the values of $C_{k,l,t} $ and the variation in the resulting observed frequencies $F_{l,t} = \frac{C_{l,t}}{N_{l,t}}$ depend on the total number of sequenced specimens, $N_{l,t}$. For example, if there were only $N_{l,t}=3$ sequences collected on a particular day in a particular location, the only possible observations of clade proportions are $F_{l,t} = {0, 1/3, 2/3, 1}$, despite predictions for individual components of $\theta_{l,t}$ being continuous on the range from 0 to 1. Therefore, when sequence counts are low, we expect the prediction intervals for the observed proportions $F_{l,t}$ to be relatively wide, whereas for large numbers of sequences the uncertainty on the observed proportions $F_{l,t}$ are expected to be narrower. Thus, properly calibrated nowcasts of the observed counts of each clade or observed frequencies of each clade would require teams to model and forecast $N_{l,t}$. To avoid this, we pre-specify that their predictions of the observed counts of each clade will be evaluated via sampling from a multinomial observation model once the number of sequences in each location and target date, $N_{l,t}$, is known so we compare the model predicted $\hat{C}_{l,t} \sim \text{Multinomial}(N_{l,t}, \hat\theta_{l,t})$ to the observed $C_{l,t}$. We allow teams to submit both point estimates of the latent clade proportions $\hat{\theta}_{l,t}$ and/or probabilistic estimates, in the form of trajectories representing draws from the joint predictive distribution, of $\hat{\theta}$ across time points and locations. 
Further details of the evaluation process are described in Section \ref{sec:modelevaluation}.  

\subsubsection{Prediction horizon}
\label{sec:predictionhorizon}
Genomic sequences tend to be reported weeks after being collected, with delays from collection to submission of sequences being highly variable. However, since we are interested in understanding the levels of circulating clades among SARS-CoV-2 infections, we index the count of sequences by the ``primary event" (the date of specimen collection) rather than the ``secondary event" (the report date), as the primary event occurs during an individual's infection. As a result, the primary event data is right-truncated, resulting in a high degree of ``backfilling'', where sequences collected at a time in the past are not reported for a significant time into the future. For this reason, the Hub collects hindcasts and nowcasts (predictions for data prior to and including the current time, which may or may not have partial observations) and short-term forecasts (predictions into the future). Counting the Wednesday nowcast date as a prediction horizon of zero, we collect daily-level predictions for 10 days into the future (the Saturday that ends the epidemic week after the Wednesday submission) and 31 days into the past (the Sunday that starts the epidemic week four weeks prior to the Wednesday submission date). Overall, six weeks (42 days) of predicted values are solicited each week. For simplicity, we will refer to the predictions (of latent clade proportions or expected observed proportions) as nowcasts for all horizons ranging from -31 to 10.

\subsubsection{Probabilistic nowcast targets}
\label{sec:probabilisticnowcasttargets}
Probabilistic nowcasts are submitted in the form of 100 samples $\hat{\theta}^{(1)}, \ldots, \hat{\theta}^{(100)}$ from the predictive distribution for $\theta$, where $\hat{\theta}^{(s)}$ is a draw from the joint predictive distribution across all time points and locations for which modelers submit. This allows for, but does not require, multivariate dependence across time and locations. Upon forecast submission, automated data validation enforces that each sample  $\hat{\theta}_{l,t}^{(s)}$  is a probability vector of length $K$-- meaning all entries are non-negative and sum to one-- at every time point and location. 

\subsubsection{Point nowcast targets}
\label{sec:pointnowcasttargets}
Teams may submit only a point forecast, only probabilistic samples, or both types of model output. Submitted point forecasts represent the mean value of $\theta$, their estimate of the underlying clade proportion for each location and target date. If a team only submits probabilistic samples, we compute the mean value for each target date and location in order to evaluate the point nowcast. 

\section{Overview of models}
\label{sec:overviewofmodels}
Nowcasts from five independently run models were submitted during the period of initial assessment (October 9th, 2024 to June 4th, 2025). Two of them were contributed by groups of academics (UMass and UGA), one by the Los Alamos National Lab (LANL), and two by the California Department of Public Health (CADPH). The University of Massachusetts (UMass) and University of Georgia (UGA) models are trained on only the GenBank data, while the LANL model is trained on data from the Global Initiative on Sharing all Influenza Data (GISAID) \parencite{Shu2017} from the U.S.and the United Kingdom. The CADPH models use California COVIDNet data, which consists of submissions from over 50 laboratory partners in California \parencite{Wadford2023}. Sequences in California COVIDNet with greater than 83 \% reference genome coverage are then submitted to the NCBI GenBank database and GISAID. Most models are a single individual model. Only the LANL model is an ensemble of multiple transformer model outputs.

The Hub creates an additional model: the Hub baseline. The Hub baseline is a Bayesian multinomial logistic regression model that is estimated using data that is fully pooled across the U.S., estimating identical samples of the predictive distribution for each jurisdiction. The model is trained using 50 days of training data prior to the target date.

Beginning in July 2025, the \code{blab-open\_hier\_mlr} and \code{blab-gisaid\_hier\_mlr} began submitting models and a Hub-ensemble model was produced and submitted.  However, these models are not included in the evaluation analysis presented here because their submissions start after the period of initial assessment (October 2024 through June 2025). 

More information on each of the models in this analysis are provided in \autoref{table:models}. Additional metadata on all models that are currently or have previously submitted to the Hub as of the writing of this manuscript can be found in the supplement (Table S1. Additional model metadata) and is regularly updated as new models submit at: \url{https://github.com/reichlab/variant-nowcast-hub/tree/main/model-metadata}. 

\afterpage{%
    \clearpage
    \thispagestyle{empty}
    \begin{landscape}
        \centering 
        \begin{longtable}{p{4cm}p{6cm}p{5.5cm}p{1.5cm}p{1.5cm}p{2cm}p{1.5cm}}
        \caption{Table 1: Models in the SARS-CoV-2 Variant Nowcast Hub during the period of initial assessment (October 9th, 2024 to June 4th, 2025). ~   \label{table:models} } \\
        \toprule
        \textbf{Model Abbreviation} & \textbf{Short Description} & \textbf{Citation} & \textbf{Data Sources} & \textbf{Locations} & \textbf{Output Type} & \textbf{Ensemble?} \\
        \midrule
        \endfirsthead
        
        \multicolumn{7}{c}%
        {{\tablename\ \thetable{} -- continued from previous page}} \\
        \toprule
        \textbf{Model Abbreviation} & \textbf{Short Description} & \textbf{Citation} & \textbf{Data Sources} & \textbf{Locations} & \textbf{Output Type} & \textbf{Ensemble?} \\
        \midrule
        \endhead
        
        \code{Hub-baseline} & A Bayesian multinomial logistic regression model that makes predictions at the national level, fit to the last 50 days of historical data. This model uses a linear in logit space model for the growth of the variants and makes the same predictions for each state. & \url{https://github.com/reichlab/variant-nowcast-model-dev-retro} &  GenBank data & All & Point and probabilistic & No \\
        \midrule
        \code{UMass-HMLR} & A Bayesian hierarchical multinomial logistic regression (HMLR) model for nowcasting COVID variants. Regression coefficents are modeled hierarchically across variants and locations. & \url{https://github.com/reichlab/variant-nowcast-model-dev-retro} & GenBank data & All & Point and probabilistic & No \\
        \midrule
        \code{UGA-multicast} & A multinomial logistic regression model with shared variant growth-rate parameters and location-specific prevalence levels, with uncertainty derived via simulation from the underlying regression coefficient uncertainty. & \url{https://github.com/sjfox/multicast/tree/main} & GenBank & All & Probabilistic & No \\
        \midrule        
        \code{LANL-CovTransformer} & A streamlined single-layer transformer architecture augmented with linear input and output layers, using embedding dimensions of 8 and dual attention heads. CovTransformer is an integrated ensemble of models, which first makes a 14 day prediction using 5 models (Stage 1 models). & Yinan Feng, Emma E Goldberg, Michael Kupperman, Xitong Zhang, Youzuo Lin, Ruian Ke, CovTransformer: A transformer model for SARS-CoV-2 lineage frequency forecasting, Virus Evolution, Volume 10, Issue 1, 2024, veae086, https://doi.org/10.1093/ve/veae086 & GenBank for inference, GISAID data from the U.S. and the U.K. for training & All & Point & No \\
        \midrule
        \code{CADPH-CATaMaran} &  A multinomial logistic regression model targeting variant observations from the past 3 months or from first observation. Sequences reported in the 30 days before the nowcast date were truncated to reduce the potential for data delays to bias the nowcast. & Nicholas G. Davies et al., Estimated transmissibility and impact of SARS-CoV-2 lineage B.1.1.7 in England. Science 372, eabg3055(2021).DOI:10.1126/science.abg3055 & California COVIDNet & California only & Point & No \\
        \midrule
        \code{CADPH-CATaLog} & This model fits a logistic growth model (binomial regression) to each target variant proportion over the past 3 months or from first observation normalizing outputs. Sequences reported in the 30 days before the nowcast date were truncated to reduce the potential for data delays to bias the nowcast. & Reichmuth ML, Hodcroft EB, Althaus CL (2023) Importation of Alpha and Delta variants during the SARS-CoV-2 epidemic in Switzerland: Phylogenetic analysis and intervention scenarios. PLOS Pathogens 19(8): e1011553. https://doi.org/10.1371/journal.ppat.1011553  & California COVIDNet & Calfornia only & Point & No \\
        \end{longtable}
        
    \end{landscape}
    \clearpage
}

\section{Model evaluation}
\label{sec:modelevaluation}
Due to the multivariate nature of the forecasts requested as well as the unique features of the data (lots of backfill and dynamic prediction targets), model evaluation is not straightforward; a few of the challenges are listed below.  

\subsection{Probabilistic forecast evaluation via multinomial sampling to generated predicted observations}
\label{sec:probabilisticforecasteval}
As was described earlier, teams submit samples $\hat{\theta}^{(1)}, \ldots, \hat{\theta}^{(100)}$ from a predictive distribution for the true proportions $\theta$ of all SARS-CoV-2 infections at the population level across locations and specimen collection date (See Section \ref{sec:forecasttarget} for more details). However, we cannot directly evaluate these predictions because we do not observe the value of $\theta_{l,t}$, instead we observe the counts of the number of sequences of each clade $k$, $C_{k, l,t}$ at each time point and location. The Hub uses the total number of sequences collected at each time point and location, $N_{l,t}$, from the evaluation dataset, and the clade proportion nowcast vectors $\hat{\theta}^{(s)}_{l,t}$ to generate nowcasts for observed clade counts, $\hat{C}_{l,t}$, by sampling from the pre-specified observation model, which is a multinomial distribution. For each sample $s$ of the probabilistic nowcast targets, $\hat{\theta}^{(s)}_{l,t}$ the Hub generates 100 predictions for observed clade counts $\hat{C}^{(1)}_{l,t}, \ldots, \hat{C}^{(100)}_{l,t}$ where each draw of $\hat{C}^{(i)}_{l,t}$ is drawn from a $\text{multinomial}(N_{l,t}, \hat{\theta}^{(s)}_{l,t})$ distribution. This results in 10,000 total draws of predicted clade count vectors to evaluate against the observed counts for each clade, day, and location.

The use of a multinomial distribution assumes that, conditional on the mean prevalence, clade assignments for the sequenced samples are independent and have probability of being in each clade equal to the population probabilities $\theta_{l,t}$. This is a plausible assumption when the number of sequenced samples, $N_{l,t}$, is small relative to the total infected population (as is generally the case in the U.S.). Furthermore, while getting around the need for teams the number of final sequences at a given collection date, the use of a multinomial distribution with size $N_{l,t}$ also introduces a specific assumption about the variation in the observation process that takes probabilities and a size $N_{l,t}$ and turns them into expected observed counts of sequences of each clade. If a team does not believe these assumptions, they may wish to modify their distribution for $\theta$ accordingly. For example, if a team believes that an overdispersed Dirichlet-Multinomial distribution would more accurately model the variation in future observations, they should add dispersion to their distribution for $\theta$. Or if a team believes that sampling is biased and some clades are underrepresented in the reported data, they may wish to modify their estimate of $\theta$ to reflect the reporting process.

These count forecasts $\hat{C}^{(1)}_{l,t}, \ldots, \hat{C}^{(10,000)}_{l,t}$ are scored on the observed counts $C_{l,t}$, using the energy score \parencite{gneiting2008assessing, jordan2019evaluating}, a proper scoring rule for multivariate data. The energy score compares the actual forecast distribution to the observed data and can be estimated in a Monte Carlo fashion using samples from the forecast distribution $\hat{C}_{l,t}$. We note that because the energy score is computed on counts, it scales with the number of sequences (in the way that weighted interval scores and continuous ranked probability scores scores also scale with counts). A smaller energy score indicates better predictive skill. 

On the Hub's GitHub, we also compute the categorical Brier score on each of the probabilistic trajectories and summarize it using the mean Brier score at each nowcast date, target date, and location. However, this metric is not used in this analysis. 

\subsection{Point forecast evaluation}
\label{sec:pointforecasteval}
Point predictions $\hat{\theta}$ are scored directly using the categorical Brier score. We use the derived formula for computing the categorical Brier score from observed clade counts and estimated clade proportion demonstrated by Susswein et al. \parencite{Susswein2023}. The categorical Brier score ranges from 0 to 1, with 0 indicating a perfect forecast and 1 indicating the worst possible forecast; thus, lower Brier scores are better. 

\subsection{Analysis of scores}
\label{sec:analysisofscores}
In order to ensure the evaluation procedure does not incentivize assigning the clade proportion to the observed clade frequency available as of the nowcast date, we only evaluate days with no observations available as of the nowcast date. The implication of this decision is that each location and nowcast date will have a different number of evaluation days, and likely a different distribution of horizon days (with locations with more and timelier sequencing being more likely to have partial observations at horizons within the nowcast period compared to locations with less data). In practice, only 9.8 \% of potential days on which we could have evaluated predictions had sequences that were already reported as of the nowcast date due to delays from collection to submission. See Fig. S2 \& S3 for plots of the proportion of days with partial observations by horizon, overall (Fig. S2), and for each individual location (Fig. S3). For completeness, we also conduct a secondary evaluation analysis using all possible evaluation days for all horizons, including days when sequences were reported at the time of the nowcast (Fig. S4, S5, S6). 

Because two of the models were only submitted for California, we analyze the state of California separately from the other 51 U.S. jurisdictions, to ensure that in both sets of locations analyzed models submitted for the majority of nowcast dates and locations. 

We summarize scores across locations and nowcast dates, and also present scores stratified by location, nowcast date, and prediction horizon.

To ensure a fair comparison between models, we compute the scaled relative skill score for each stratification. The relative skill score of a model is the geometric mean of the pairwise score ratios of all pairs containing that model. The score ratio is computed using only the set of overlapping nowcasts for a particular pair of models, within a specified strata (e.g. location or nowcast date). The scaled relative skill score (SRSS) is computed by dividing the relative skill score of a model by the relative skill score of the baseline model \parencite{scoringutils}. In our case, we use the Hub-baseline as the baseline model. The scaled relative skill is computed using both the energy and Brier score, as the Brier score calculation allows us to compare more models because we can evaluate models that submit only the mean, whereas the energy score allows us to assess the full distribution of estimates.

When we summarize either the Brier score or the energy score overall, by horizon, by nowcast date, and by location, we compute the arithmetic mean of the score for each target date, nowcast date, and location. This means that in some instances, depending on the stratification, we may be comparing scores across a different set of nowcasts, since not all models submitted for all nowcast dates and locations across all weeks, and each nowcast has a different set of horizons that are evaluated. Of note, the energy score scales with the number of counts being evaluated, which means that when there are a larger number of sequences, we expect a higher energy score. We chose to include sequence volumes alongside our figures by nowcast date and location so that the interpretation of energy score could be done within the context of the number of evaluation sequences. 

For horizons and jurisdictions with no observed sequences during the evaluation period, no scores are computed and these days are excluded from the analysis. 

\subsection{AI disclosure}
Generative AI tools were used to assist with preparation of the manuscript text and as part of code review for both the Variant Nowcast Hub GitHub and the analysis evaluating the performance of submitted variant nowcasts. Coderabbitai https://github.com/apps/coderabbitai was used for automated code review. Claude Pro version 4 https://claude.ai/ was used for assistance with writing code and for revisions of text in the manuscript. 

\section{Results}
\subsection{The data: SARS-CoV-2 sequence counts and observed clade frequencies across U.S. jurisdictions}

\subsubsection{SARS-CoV-2 variant dynamics in the U.S. using the final available data}

\begin{figure}[h]
    \centering
    \includegraphics[width=0.8\textwidth]{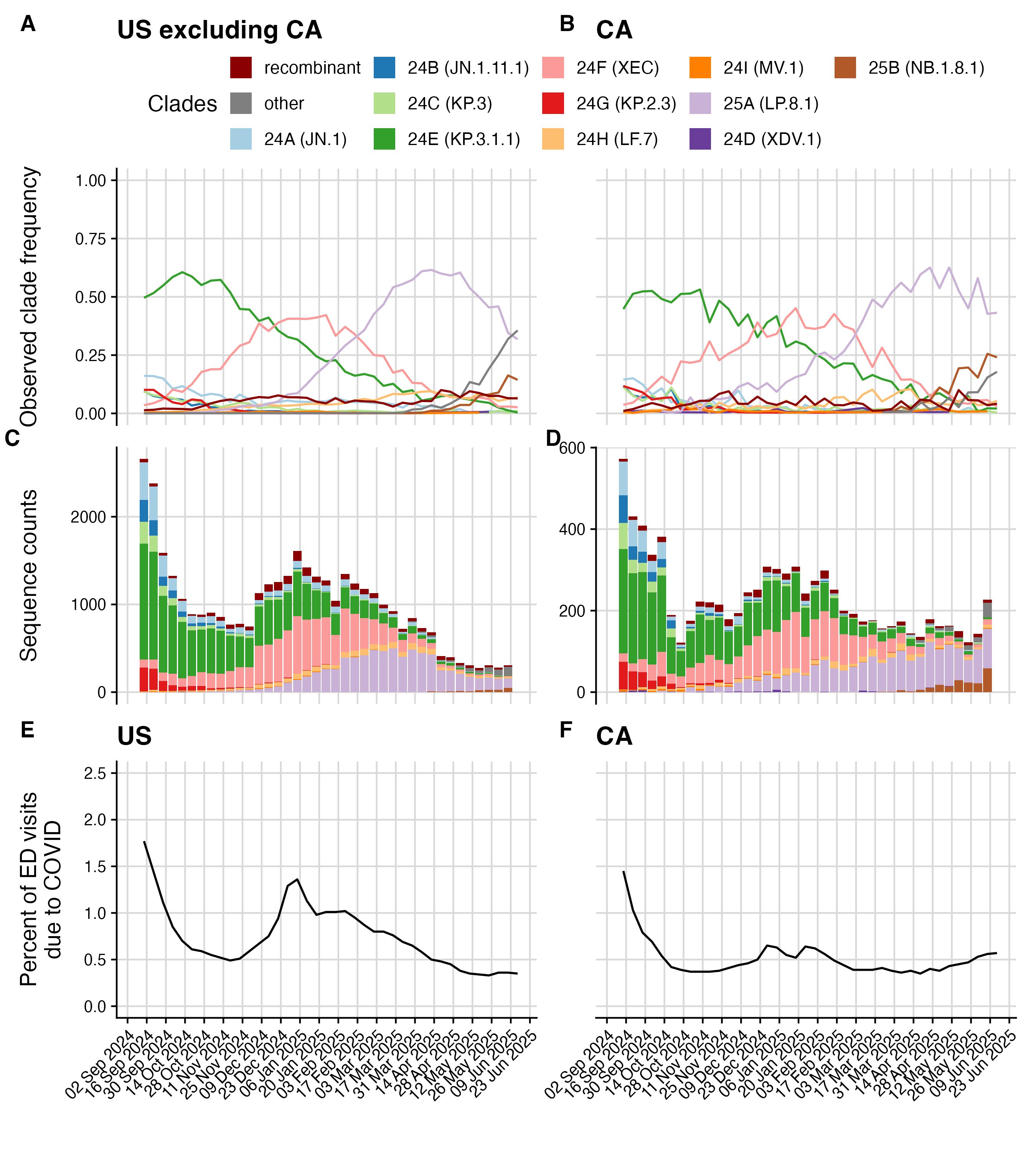}
    \caption{Weekly SARS-CoV-2 variant dynamics in the U.S. from September 2nd, 2024 to June 14th, 2025 using the final set of sequences available 90 days after the last nowcast date. Data from all of the U.S. states, D.C. and Puerto Rico excluding California is on the left and data from California is on the right, for consistency with the summaries of model performance. (A \& B). Final observed weekly clade frequencies. (C \& D). Number of weekly sequences collected colored by clade. (E \& F). Weekly percent of emergency department (ED) visits due to COVID-19 in the U.S. and California (right).}
    \label{fig:USvariantfreq}
\end{figure}

In both California (right) and the rest of the U.S. (left), clade 24E (green) began as the most prevalent clade (around 50 percent prevalence) in September 2024 and declined in prevalence slowly (\autoref{fig:USvariantfreq}). It was first replaced by clade 24F (pink) in October and November 2024, which was then replaced by clade 25A (purple) which became the dominant lineage at the beginning of March 2025. Clade 25A then began to be replaced by clade 25B (brown) and others in May 2025. 

Within the U.S., the number of sequences collected from September 2nd, 2024 to June 14th, 2025 varied by state, ranging from 9,220 sequences collected in California to only 1 sequence collected from Idaho, Maine, and Nebraska (Fig. S7). Proportionally more sequences come from large jurisdictions than would be expected based on relative population size alone, with 90\%  of all sequences submitted during this period coming from states that represent 49\% of the total U.S. population. Fig. S8 shows the number of sequence counts by location and nowcast date. 

Across jurisdictions, there is also some indication of differences in variant dynamics within individual states such as California, Minnesota, and Illinois (Fig. S9-11). Focusing on 25A's emergence, we see that there is some variability in the timing and speed of emergence (Fig. S12). 

 \subsubsection{Example SARS-CoV-2 sequence data available to modelers as of February 19th, 2025 versus when nowcasts are evaluated}

\begin{figure}[h]
    \centering
    \includegraphics[width=0.8\textwidth]{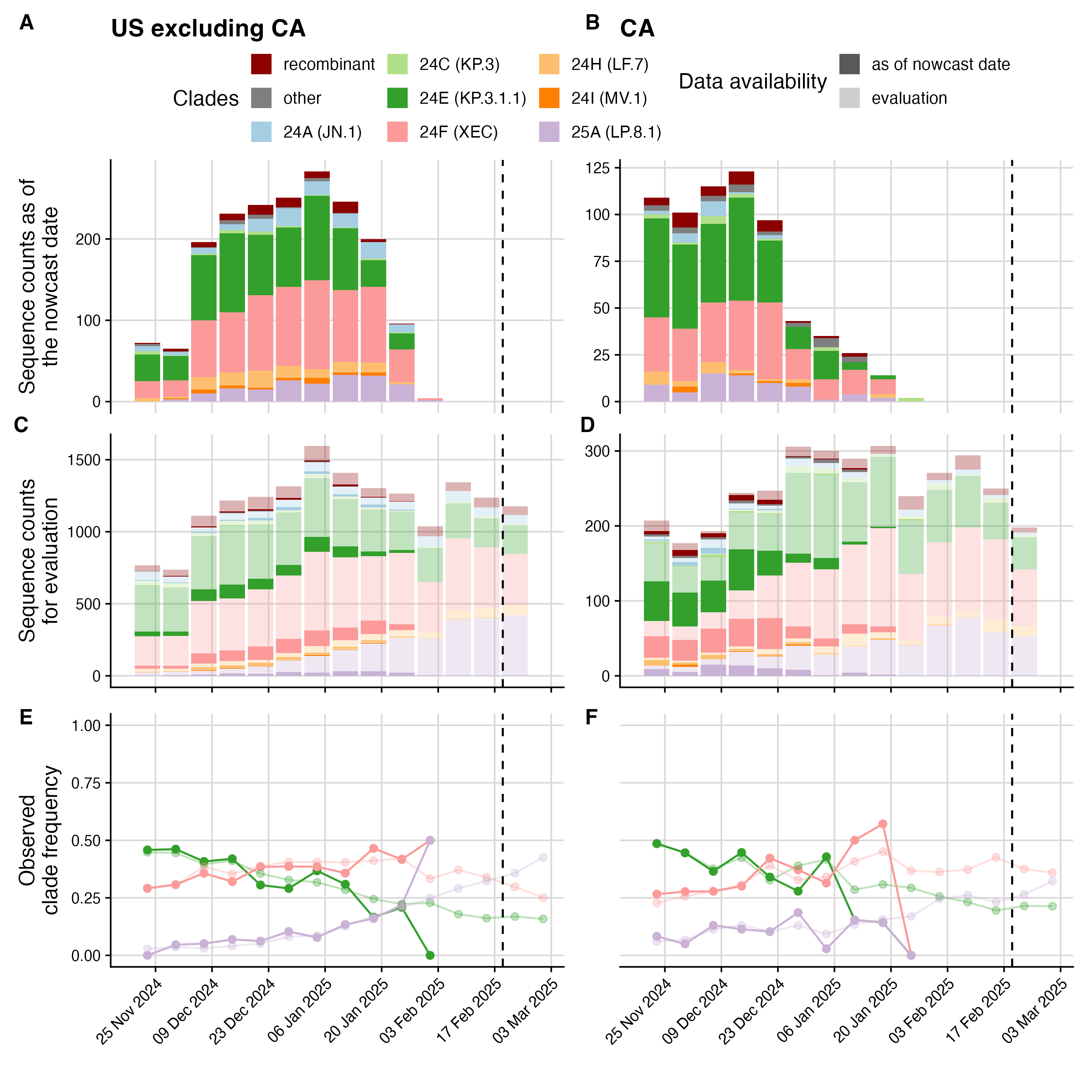}
    \caption{SARS-CoV-2 sequence counts and clade frequencies for an example nowcast date of February 19th, 2025 versus the counts and frequencies later available at the time of evaluation. (A \& B). Sequence counts as of the nowcast date colored by clade. (C \& D). Sequence counts as of the nowcast date (solid) and which are later available (shaded). (E \& F). Observed clade frequencies as of the nowcast date (solid) and at the time of evaluation (shaded) in the U.S. excluding California (left) and California (right).}
    \label{fig:dataasof}
\end{figure}

The dynamics in \autoref{fig:USvariantfreq} show clade frequencies varying somewhat smoothly, however, the observed clade frequency that modelers have available to them as of the nowcast date (\autoref{fig:dataasof}A \& B) differs significantly from the final sequences available \autoref{fig:dataasof}C \& D), with the solid lines (\autoref{fig:dataasof}E \& F) representing the observed clade frequencies as of the nowcast date versus the shaded lines representing the observed clade frequencies as of the time we evaluate the nowcasts (\autoref{fig:dataasof}E \& F).

In this case, we see that in the U.S. excluding California, the data as of the nowcast date shows a sharp increase in the prevalence of 25A from below 25\% to 50\% abundance in the last week with available data. However, in the observed frequencies at the time of evaluation, the frequency of 25A increases much more gradually over the time period of interest (\autoref{fig:dataasof}E).  In California, the last week with any observed sequences is a week earlier, and in this instance we only observe a single sequence of clade 24C, resulting in what appears to be a sharp decline in both the prevalence of 24F and 25C (\autoref{fig:dataasof}F). Again, the observed frequencies at the time of evaluation instead show a much more gradual increase in the frequency of 25C and stabilization of both 24F and 24E. 

\subsection{Model estimates and predicted observations on February 19th, 2025 in California, Illinois, and Minnesota}

To illustrate what modelers submit for each location at a particular nowcast date, we focus on the nowcasts submitted on February 19th, 2025 in three example locations in the U.S.: California, Illinois, and Minnesota. These locations were chosen as examples as they have sufficient levels of sequencing to be able to make comparisons of the expected observed frequencies at a daily scale compared to the predicted observed frequencies, but they differ enough across the time range and across locations to provide an illustration of the impact varying levels of sequencing has on the predicted observations. 

\begin{figure}[h]
    \centering
    \includegraphics[width=0.8\textwidth]{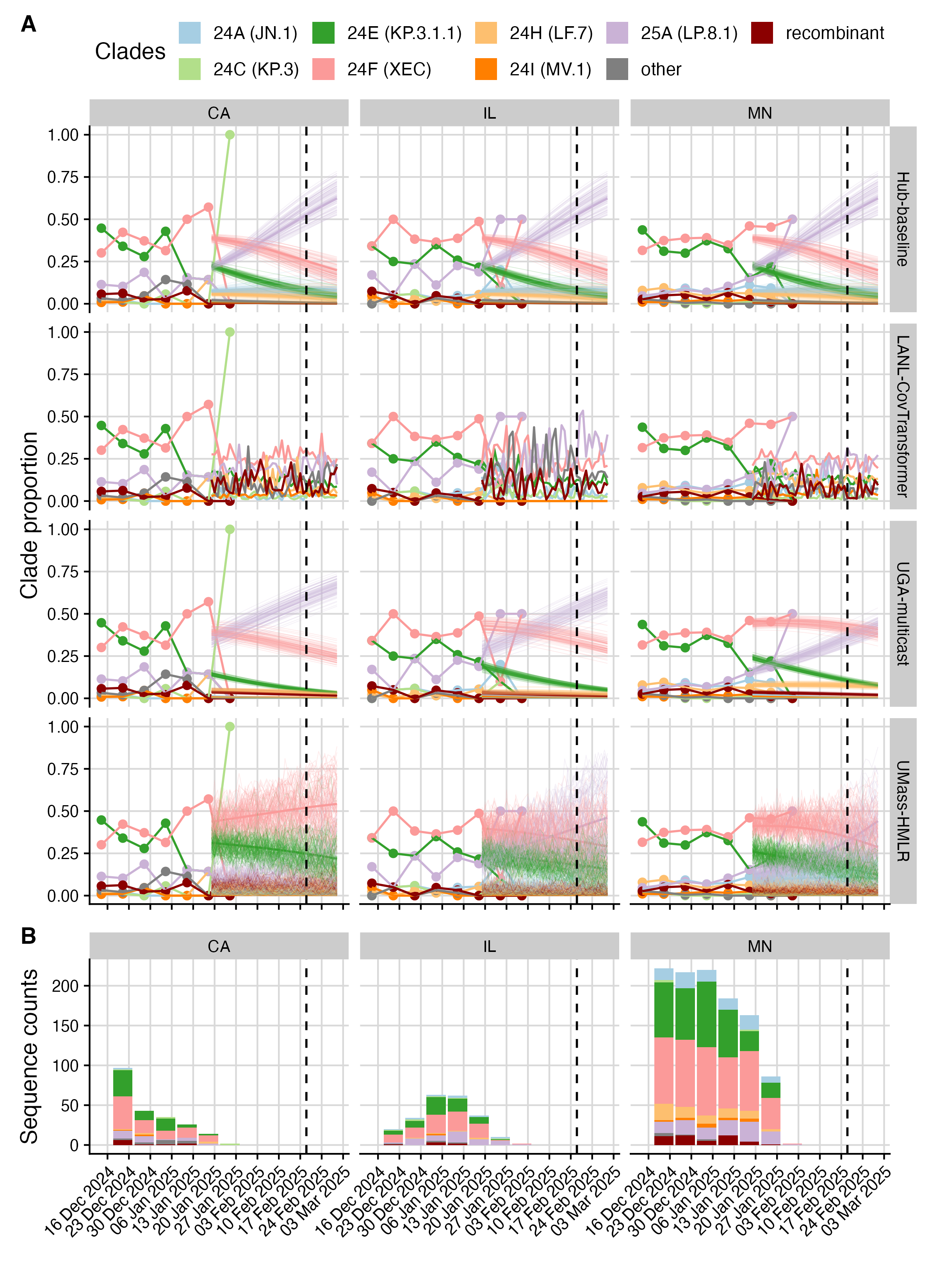}
    \caption{Example model outputs alongside the observed clade frequencies as of February 19th, 2025. (A) Nowcasted estimates of the true clade proportions alongside the observed clade frequencies for California (left), Illinois (middle) and Minnesota (right) for the Hub-baseline (top), LANL-CoVTransformer (upper middle), UGA-multicast (lower-middle) and UMass-HMLR (bottom). Colors indicate clade. Solid lines with points indicate the observed clade frequency. Model trajectories (samples from the predictive distribution for latent clade proportions) are plotted if submitted, with the mean plotted as a thicker line if submitted. (B) Number of sequences colored by clade available as of the nowcast date. The Hub-baseline and UMass-HMLR models submit both trajectories and mean predicted clade proportions, while UGA-multicast only submitted trajectories and LANL-CoVTransformer only submitted point estimates of clade proportions. The Hub-baseline is fit to all data in the U.S. and predicts the same clade proportions for all locations.}
    \label{fig:modelnowcasts}
\end{figure}

For nowcasts made on February 19th, 2025, the degree to which the observed clade frequencies available in real-time differed from the evaluation data differed across locations (\autoref{fig:modelnowcasts}).  
Specifically, California, Illinois, and Minnesota show variable numbers of sequences and corresponding observed clade frequencies at the time of the nowcast (\autoref{fig:modelnowcasts}B). 

Based on the available data, models made differing predictions about the current and future trajectory of the different clades.  For the LANL-CovTransformer, UGA-multicast, and UMass-HMLR, it is evident that the model is tuned to the specific observed dynamics in each state. For example, UMass-HMLR predicts an increase in prevalence for 25A in Illinois and Minnesota but does not predict the same increase for California (\autoref{fig:modelnowcasts}A). The CADPH-CATaLog model point estimates for California are shown in Fig. S13. CADPH-CATaMaran did not submit a nowcast on February 19th, 2025.

\begin{figure}[h]
    \centering
    \includegraphics[width=0.7\textwidth]{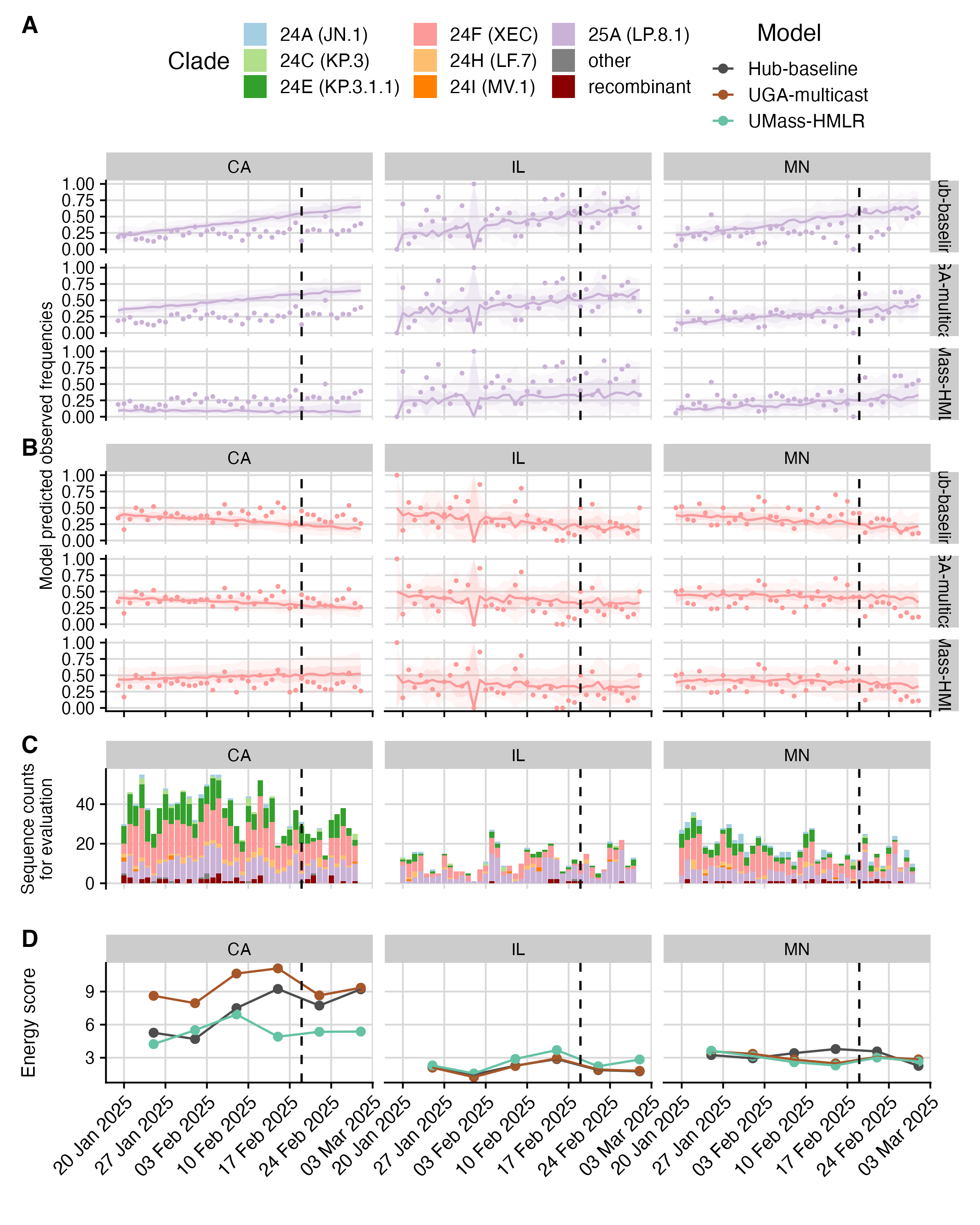}
    \caption{Example model predicted observed frequencies with prediction intervals as of February 19th, 2025. Model predicted observed clade frequencies (points) alongside the final observed clade frequency for California (left), Illinois (middle) and Minnesota (right) predicted by the Hub-baseline (top), UGA-multicast (middle) and UMass-HMLR (bottom) for clade 25A (A) and clade 24F (B). Shading indicates the central 50th and 90th percentile prediction intervals respectively, solid line indicates the median predicted observed frequency. Colors indicate clade. (C) Final number of sequences for evaluation on each target date, colored by clade. (D) Average energy score for each nowcast date for each model, colored by model. }
    \label{fig:predobs}
\end{figure}

Three models submitted probabilistic nowcasts in the form of trajectories (see Section \ref{sec:probabilisticnowcasttargets}) of latent clade proportions on February 19th, 2025.
These were turned into prediction intervals through the multinomial sampling process described in the Methods section and then compared to the final observed sequence count at each day, for all clades. 
The sequence counts for evaluation available each day (\autoref{fig:predobs}C) dictate the range of possible observed clade frequencies, which are here summarised using the central 50th and 90th prediction intervals. For example, on a day with only 1 sequence collected (e.g. Illinois on February 1st, 2025) the prediction intervals will range from 0 to 1  (\autoref{fig:predobs}A middle), because the observed clade frequency can only have a value of 0 or 1. Generally speaking, we expect wider prediction intervals for days with fewer sequences, and tighter prediction intervals for days with a higher number of sequences for evaluation.

Visually, we can see that for both the Hub-baseline and UGA-multicast, the observed frequency of clade 25A falls below the central 50th and 90th percentile prediction intervals for the majority of the dates evaluated in this nowcast date, whereas the UMass-HMLR model's prediction intervals contain more of the observed clade frequencies. 

\subsection{Submission summary}
From October 7th, 2024 to June 9th 2025, a total of 6 models were submitted to the U.S. SARS-CoV-2 Variant Nowcast Hub, including the Hub-baseline. The Hub-baseline was submitted for all locations and nowcast dates for both trajectories and mean nowcasts, though submissions before April 2nd, 2025 were submitted retrospectively, using the data available as of each nowcast date. UMass-HMLR submitted trajectories and mean nowcasts for all nowcast dates, excluding some jurisdictions on certain dates. UGA-Multicast submitted trajectories for all jurisdictions and all nowcast dates starting on October 28th, 2024. LANL-CoVTransformer submitted mean nowcasts for all  nowcast dates from October 12, 2024 onwards, excluding some jurisdictions on certain dates. CADPH-CATaLog submitted mean nowcasts for only California from October 28th, 2024 onwards excluding April 7th, 2025. CADPH-CATaMaran submitted mean forecasts for only California from March 31st, 2025 onwards excluding April 7th, 2025. See Fig. S14 for heatmaps indicating the dates and locations that each model submitted for and Fig. S15 for the number of models submitted at each nowcast date and location over the time period.

For all locations except for California, only UMass-HMLR, UGA-multicast, LANL-CovTransformer, and the Hub-baseline have scores, with LANL-CoVTransformer only submitting a point forecast (and therefore not in the energy score comparison, as only models submitting samples of the full predictive distribution are sampled assuming a multinomial observation process). For California, we have the two additional California models, CADPH-CATaLog and CADPH-CATaMaran, in the point forecast comparison of the Brier scores. 

\subsection{Summary of model performance}

\subsubsection{Overall performance across jurisdictions and nowcast dates }

\begin{figure}[h]
    \centering
    \includegraphics[width=0.8\textwidth]{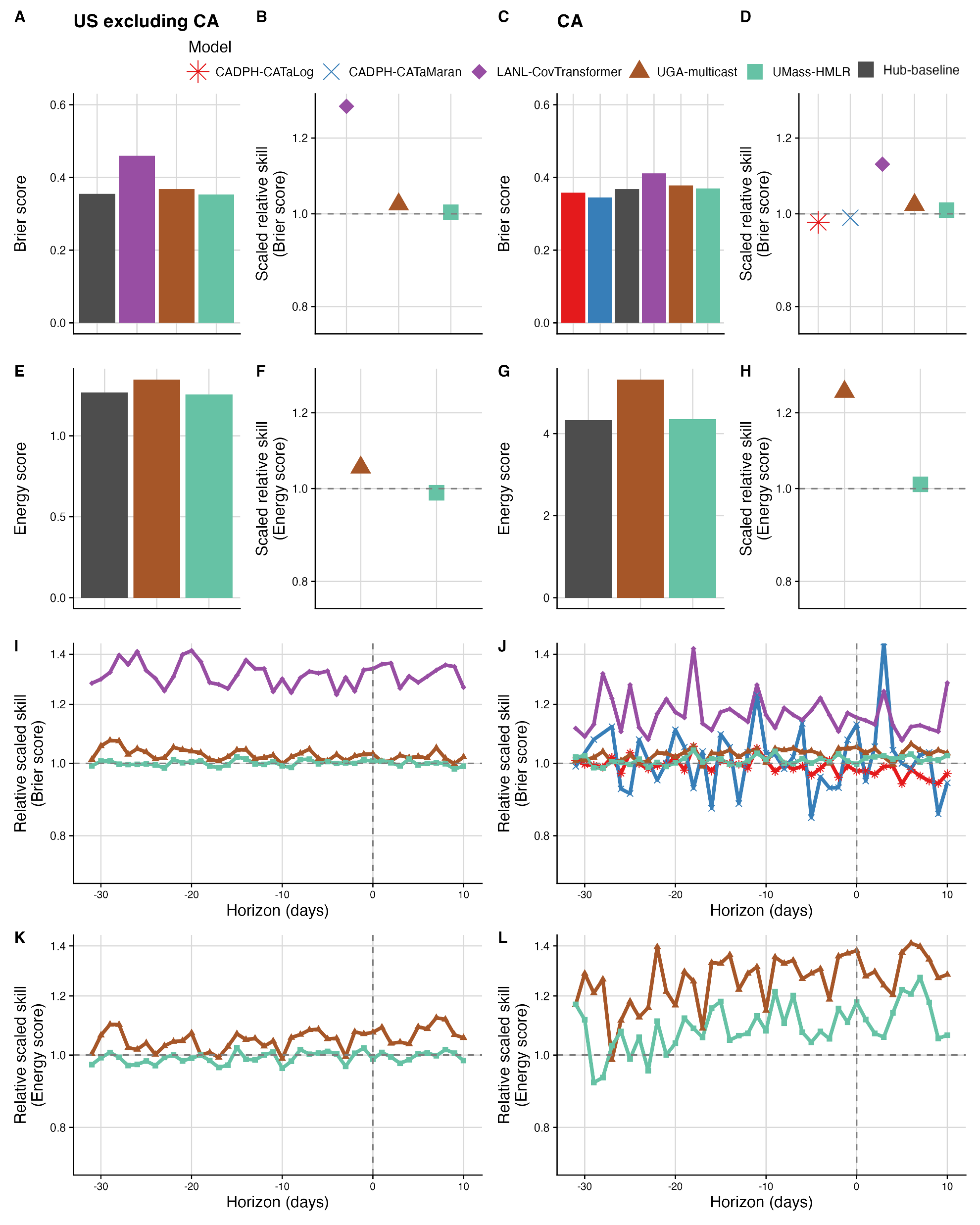}
    \caption{Score comparison over all evaluated nowcast dates, horizons, and jurisdictions (referred to as overall). Energy and Brier scores for each model are averaged across all of the nowcast dates, horizons, and jurisdictions for which the model submitted, stratified by the U.S. excluding California on the left and California only on the right. Color indicates model. (A \& C). Absolute Brier score overall. Brier scores are computed using the mean nowcast. (B \& D). Scaled relative skill on the Brier score overall. SRSS $<$ 1 indicates improved performance compared to the Hub-baseline, SRSS $>$ 1 indicates poorer performance. (E \& G). Absolute energy score overall. (F \& H). Scaled relative skill on the energy score overall. (I \& J). Scaled relative skill on the Brier score by nowcast horizon, with vertical dashed line indicating horizon 0 (the nowcast date). (K \& L). Scale relative skill on the energy score by nowcast horizon. Scores are summarized across nowcast dates. Models submitting only means are evaluated using only the Brier score on the point estimate of the mean, models submitting trajectories are evaluated using the energy score.}
    \label{fig:overallscorecomparison}
\end{figure}

In the U.S. excluding California, we find that none of the models perform better than the Hub-baseline via the Brier score on the point estimate of the clade proportion, though UMass-HMLR scores effectively equivalently (SRSS = 1.003 \autoref{fig:overallscorecomparison}A and B). When assessing the full distribution via the energy score, UMass-HLMR performs effectively equivalently to the Hub-baseline (SRSS = 0.99 \autoref{fig:overallscorecomparison}E \& F). Alternative summaries of the scaled relative skill score, in which the scaled relative skill score is computed for each horizon, nowcast date, and location, and then averaged, enforcing equal weighting regardless of the amount of evaluation data, reveals similar results, with slightly better relative performance of individual models when computed in this manner (Fig. S16 \& S17). We also visualized the distribution of scaled relative skill scores, using heatmaps and histograms, on the Brier and energy scores for each combination of nowcast date and locations which had greater than 5 sequences for evaluation, which demonstrated the heterogeneity in performance and relative symmetry in terms of improved versus reduced performance compared to the baseline for most models (Fig. S18-S21).

In California, both CADPH models perform marginally better than the baseline as measured by the Brier score (SRSS = 0.98 and 0.99 for CADPH-CATaLog and CADPH-CATaMaran respectively \autoref{fig:overallscorecomparison}C \& D). Both the UMass-HMLR and UGA-multicast perform slightly worse than the Hub-baseline when assessing the full predictive distribution via the energy score (SRSS = 1.01 for UMass-HMLR and 1.26 for UGA-Multicast \autoref{fig:overallscorecomparison}G \& H). Again, alternative summaries in which the scaled relative skill is averaged across all horizons revealed similar rankings of models, but in this case slightly reduced performance when averaging across individual days (Fig. S22 \& S23). The California specific models, CADPH-CATaLog and CADPH-CATaMaran are trained on data with an increased timeliness and magnitude of sequence data (Fig. S24 \& S25) as of the forecast date compared to the data available in the GenBank data which was the data source used by the other teams and the Hub baseline. 

Summarizing the scores by horizon across nowcast dates and locations in the U.S. excluding California, UMass-HMLR performs similar to the Hub-baseline via both the Brier score and energy score, whereas UGA-multicast performs slightly worse on average in the forecast period (horizons 0 to 10) and LANL-CovTransformer performs consistently worse than all three other models across horizons via the Brier score (\autoref{fig:overallscorecomparison}I \& K). 

In California, CADPH-CATaLog performs consistently better via the Brier score than the Hub-baseline in the late nowcast and early forecast period (horizons -5 to 10 days, \autoref{fig:overallscorecomparison}J). The scaled relative skill of CADPH-CATaMaran is highly variable, which may be due to its limited submission dates (Fig. S15). When assessing the full predictive distribution via the energy score in California both UMass-HMLR and UGA-multicast perform worse in the recent horizons days than the Hub-baseline. 

Generally all models tend to perform better at horizons further in the past (Fig. S26). Energy scores are higher when sequencing counts are higher, hence the weekday patterns observed when we look at the absolute energy across horizon days, which corresponds to higher energy scores and sequence counts on weekdays compared to weekends (Fig. S26B \& D). 

In the supplement, we show the overall scoring results without excluding the days for which there were already sequences observed within the -31 to 10 day nowcast horizons (Fig. S4). The overall findings remain the same. 

\subsubsection{Performance by jurisdiction}

\begin{figure}[h]
    \centering
    \includegraphics[width=0.8\textwidth]{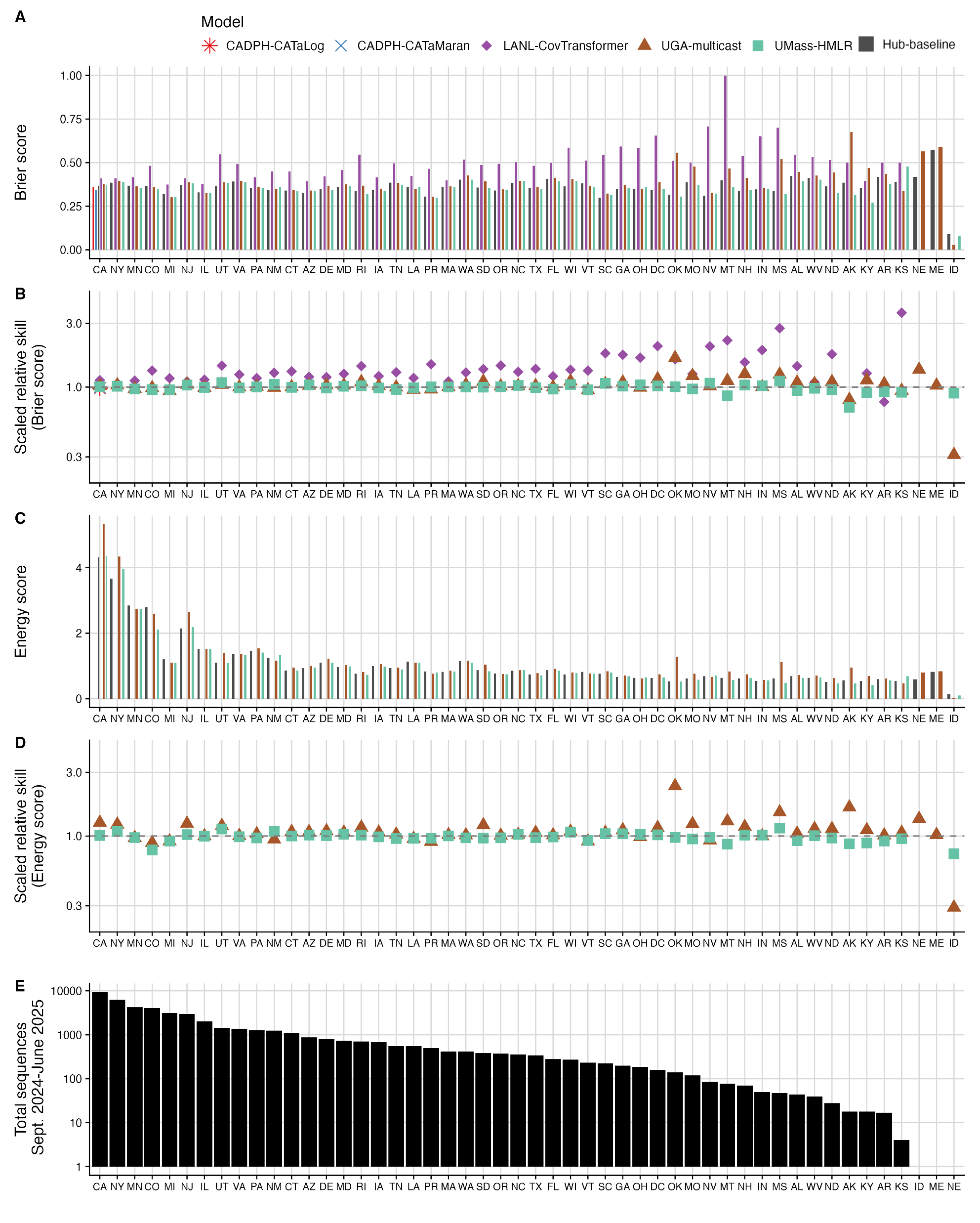}
    \caption{Score comparison of the absolute and relative Brier and energy scores across nowcast dates and horizons for each location. Jurisdictions are shown in descending order by the total number of sequences during the assessment period. Absolute Brier score averaged across nowcast dates and horizons for each jurisdiction. Brier scores are computed using the mean nowcast. B. Scaled relative skill on the Brier score versus the Hub-baseline. SRSS $<$ 1 indicates improved performance over the Hub-baseline, SRSS $>$ 1 indicates reduced performance. C. Absolute energy score averaged across nowcast dates and target dates for each jurisdiction. Energy scores are computed using the full predictive distribution submitted. D. Scaled relative skill on the Energy score versus the Hub-baseline. E. Total number of sequences collected and submitted by the time of the final evaluation dataset for each jurisdiction, ordered by the total number. Y-axis is in log scale. Color indicates model.}
    \label{fig:scoresbylocation}
\end{figure}

As described in Fig. S26,  absolute energy scores tend to be higher for all models when more sequences are available in states like California, New York, Minnesota, and Colorado (\autoref{fig:scoresbylocation} A \& E). In general, for both the relative Brier and energy score we see that individual models score more similarly to the Hub-baseline for jurisdictions with higher sequence counts. Model performance is more variable between models in jurisdictions with low sequence counts. The scaled relative skill score on the Brier score shows the LANL-CovTransformer appears to generally perform worse as sequence counts decline, whereas UGA-multicast and UMass-HMLR are more stable across jurisdictions in comparison (\autoref{fig:scoresbylocation}A). 

For certain jurisdictions, both UGA-multicast and UMass-HMLR perform better than the Hub-baseline by scaled relative skill on the energy score; examples include Colorado (SRSS of 0.79 and 0.89 for UMass-HMLR and UGA-multicast), Puerto Rico (SRSS of 0.96 and 0.91 for UMass-HMLR and UGA-multicast),  and Vermont (SRSS of 0.93 and 0.91 for UMass-HMLR and UGA-multicast) (\autoref{fig:scoresbylocation}D). However, there are also states where both models perform worse e.g. in New York (SRSS of 1.08 and 1.22 or UMass-HMLR and UGA-multicast), Utah (SRSS of 1.12 and 1.19 for UMass-HMLR and UGA-multicast), and Georgia (SRSS of 1.04 and 1.11 for UMass-HMLR and UGA-multicast). 

In the supplement, we show the scoring results by jurisdiction without excluding the days for which there were already sequences observed within the -31 to 10 day nowcast horizons (Fig. S5). The overall findings remain the same. 

\subsubsection{Performance by nowcast date}

\begin{figure}[h]
    \centering
    \includegraphics[width=0.8\textwidth]{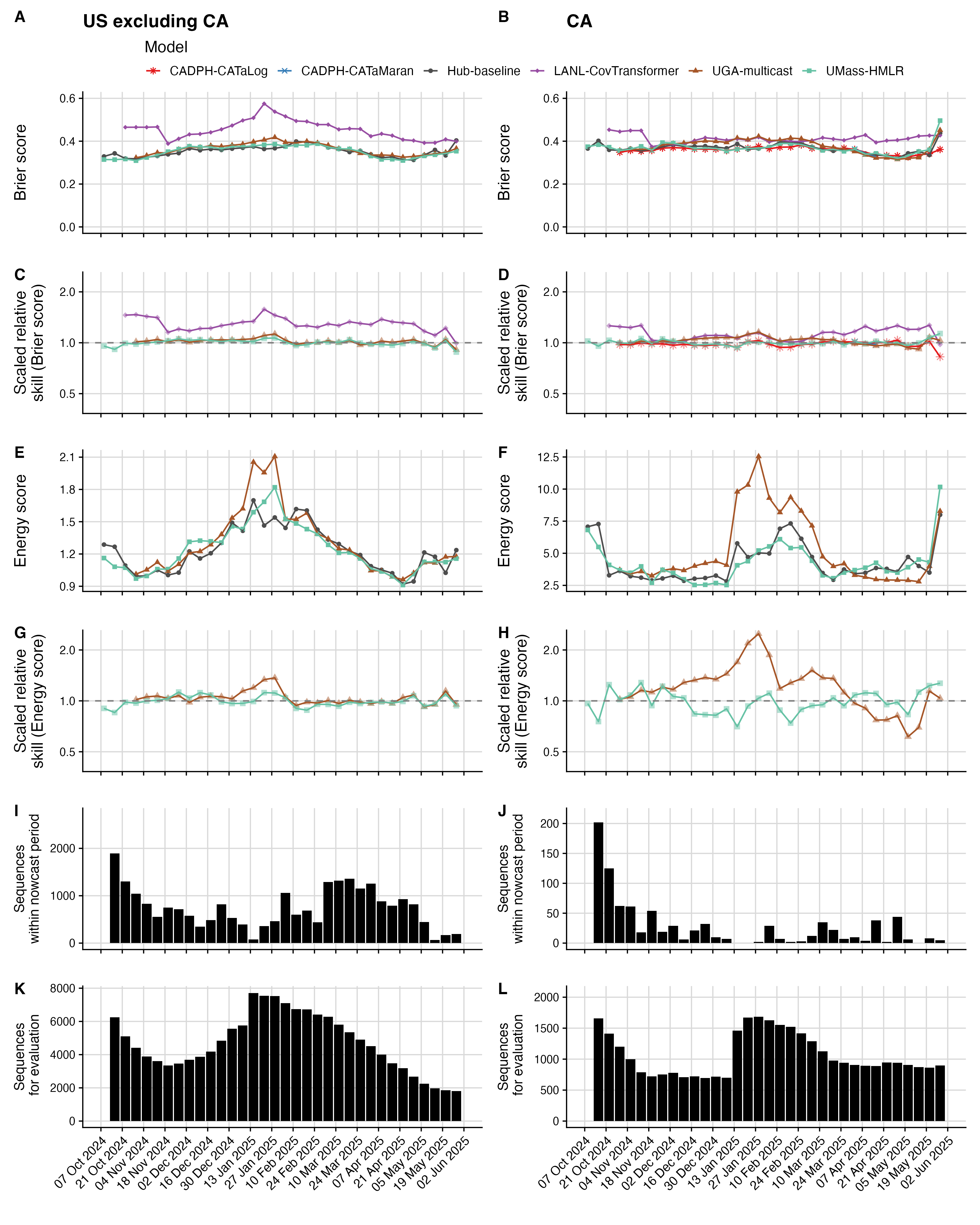}
    \caption{Score comparison by nowcast date stratified by the U.S. excluding California (left) and California (right). (A \& B). Absolute Brier score at each nowcast date. (C \& D). Scaled relative skill on the Brier score at each nowcast date. RSS $<$ 1 indicates improved performance over the Hub-baseline, RSS $>$ 1 indicates reduced performance. (E \& F). Absolute energy score at each nowcast date. (G \& H). Scaled relative skill on the energy score at each nowcast date. (I \& J). Number of sequences available within the nowcast period. (K \& L). Number of sequences for evaluation by nowcast date.}
    \label{fig:scoresbynowcastdate}
\end{figure}

Across time, all models had somewhat similar trends in Brier and energy scores in both the U.S. excluding California and California (\autoref{fig:scoresbynowcastdate}A, B, E, \& F). 

In both the U.S. excluding California and California, the scaled relative skill for both the Brier score and energy scores shows an increase in mid-to-late January of 2025 for UMass-HMLR and UGA-multicast (\autoref{fig:scoresbynowcastdate}C \& G); otherwise, these models perform similarly to the Hub-baseline. The LANL-CovTransformer performs consistently worse than the Hub-baseline across nowcast dates in the U.S. excluding California. 

In California, scaled relative skill scores on the Brier score are more similar to the Hub-baseline across models, with a few nowcast dates where CADPH-CATaLog performs better (\autoref{fig:scoresbynowcastdate}D). The scaled relative skill on the energy score shows more of a difference between models over time (\autoref{fig:scoresbynowcastdate}H). Notably, sequence counts available as of the nowcast during the period from mid December to June 2025 are lower than was observed in early October of 2024 for CA (\autoref{fig:scoresbynowcastdate}J), which may have an impact on the variability in scores across individual models. 

In the supplement, we show the scoring results by nowcast date without excluding the days for which there were already sequences observed within the -31 to 10 day nowcast horizons (Fig. S6). The overall findings remain the same. 

\subsubsection{Performance during variant emergence: focus on 25A in three example jurisdictions}

\begin{figure}[h]
    \centering
    \includegraphics[width=0.6\textwidth]{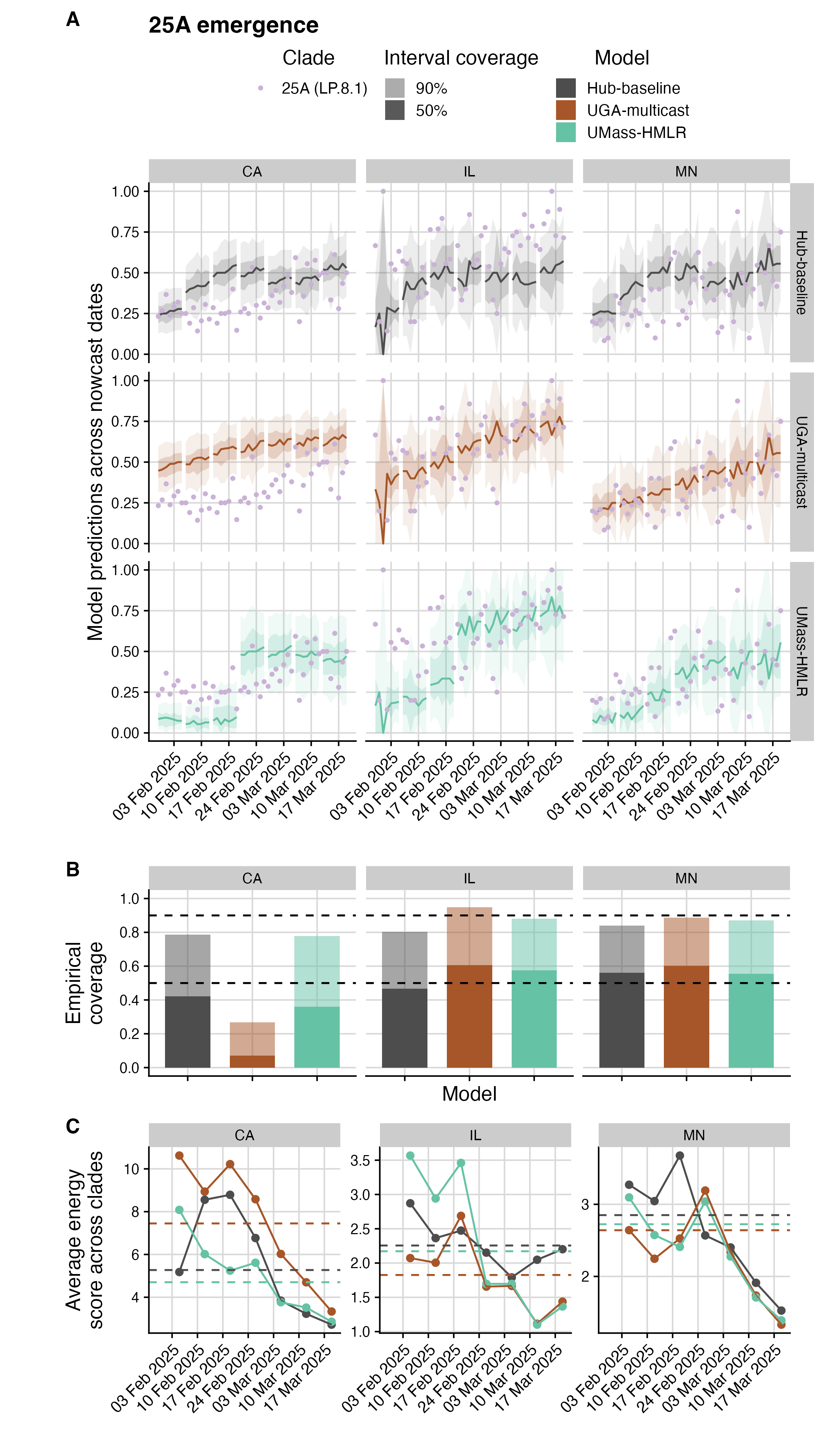}
    \caption{Visual and quantitative performance comparison of probabilistic models during emergence of 25A in California (left), Illinois (middle) and Minnesota (right). Colors indicate model. A. Visual comparison of model predicted observed proportions of 25A for a week behind nowcast (horizons -6 to 0). Shading indicates daily prediction intervals, with solid line indicating the median estimated proportion. Purple points indicate the observed 25A proportion on each day.  Variation in size of prediction intervals across time is due to variation in the final number of observed sequences. B. Empirical versus nominal coverage  of observed 25A proportions averaged across the time frame of 25A emergence for all nowcast and forecast horizons (-31 to 10). Dark shading indicates 50\% prediction intervals, light shading indicates 90\% prediction intervals. Horizontal dashed lines at 50\% and 90\% indicated nominal coverage. Well-calibrated forecasts will have empirical coverage close to the nominal coverage (e.g. 50\% of observed proportions will fall within the 50\% prediction interval). C. Average energy scores across clades for each model by nowcast date. Horizontal dashed lines indicate the average energy score across all nowcast dates during the time period of 25A emergence.}
    \label{fig:zoom25A}
\end{figure}

By focusing on forecasts made during the increase in prevalence of 25A from February to the end of March, 2025, we see that models tended to perform worse early in the clade's emergence and perform better once the clade rose to a significant abundance (\autoref{fig:zoom25A}). Clade 25A became a clade to monitor on February 5th, 2025. Prior to that, sequences were either being designated as a different clade (likely 24B, which is its closest descendant), and/or the number of sequences labeled as 25A was insufficient to warrant it being among one of the modeled clades per our clade selection algorithm (see  Section \ref{sec:cladelist} for more details). We first visually compare the nowcasted model predicted daily observed proportions from the Hub-baseline, UGA-multicast, and UMass-HMLR models for horizon days -6 to 0 (a hindcast and a nowcast), for California, Illinois, and Minnesota (\autoref{fig:zoom25A}A). The proportion of observed clade proportions falling within the 50 \% and 90 \% prediction intervals is summarized in \autoref{fig:zoom25A}B across all nowcast and forecast horizons (-31 to 10). We observe that in California, all models had lower coverages compared to their nominal values, with UGA-multicast having the lowest proportion of observed data within its 50\% and 90\% prediction intervals. Coverage was generally better in Illinois and Minnesota across all models.

Investigating overall energy scores for each model across nowcast dates (\autoref{fig:zoom25A}C), we see that broadly models begin to perform better (decreasing energy scores) across the period of time during which 25A comes to dominate, though the direction of the bias doesn't appear to follow a consistent pattern (Fig. S27). There did not appear to be a discernible pattern between which models perform best during this period across the three jurisdictions examined. 

\section{Discussion}
The United States SARS-CoV-2 Variant Nowcast Hub began operations on October 9th, 2024, and as of this writing includes seven regularly submitting models alongside a Hub baseline and an ensemble model. During the initial assessment period (October 9th, 2024 to June 4th, 2025), which spanned the decline of clade 24E, emergence of 24F, and eventual takeover by 25A, we evaluated two probabilistic models, three point nowcast models, and the Hub baseline. During this period, most submitted models performed comparably to or slightly worse than the Hub baseline. Of note, the Hub baseline model differs from other forecasting Hub baselines \parencite{FluSightHub, cramer2022evaluation}, which typically use a persistence model-- carrying through the most recent observation into the future. Here, the Hub baseline is a Bayesian multinomial logistic regression model fit to all U.S. sequencing data, generating identical predictions for all jurisdictions. We found that models submitted by teams focusing exclusively on California showed modest improvements over the baseline, likely due to more abundant and timely in-state sequencing data (Fig. S24  \& S25). The baseline's strong performance likely reflects an advantage of "fully pooling" all sequencing data in the U.S., as there was a substantial decline in nationwide sequencing since the pandemic peak, suggesting that current data sparsity limits advantages of modeling approaches which allow for greater differences in trends across jurisdictions. Of note, these findings are not generalisable to future/current time points, as we see that more recent nowcasts (\url{https://reichlab.io/variant-nowcast-hub-dashboard/explore.html}) show a reduction in performance of the baseline, perhaps due to the model overreacting to population-level "outlier" data. Additionally, had we used a persistence model as a baseline (a model which carries the last observed data point into the nowcast/forecast horizon), we expect that results relative to baseline would have been difficult to interpret due to poor baseline performance as a result of low levels of sequencing at recent dates (because the observed frequency is a function of the number of sequences, limiting the observed proportions at small numbers of sequences). One reason for reduced performance in the LANL-CovTransformer may be that it was originally designed and trained to make intermediate term forecasts (e.g. more than 14 days ahead) rather than nowcasts or short-term forecasts.  However, models remain most valuable precisely when data are sparse and observed proportions are noisy-- conditions under which relying on raw observations alone yields delayed or misleading trends.

This work makes two key methodological advances for evaluating variant nowcasts in real time. First, we address the challenge of changing clade assignments over time by using time-stamped clade assignment models from Nextclade \parencite{Hadfield2018} accessed via CladeTime \parencite{cladetime}, enabling retrospective assignment of sequences to clades as they would have been designated at the nowcast date. This approach allows evaluation within the set of clades actually available to modelers at the time of the nowcast, addressing a limitation in prior fitness model evaluations \parencite{Susswein2023, abousamra2024fitness} which either were limited to a window where clade assignments were largely constant or reran model predictions retrospectively to obtain comparable clade assignments. Second, we implement a framework for evaluating predictive distributions of latent proportions when the observation, counts of a particular clade, depend on a nuisance parameter (final number of sequences). This uses an observation model to generate expected predicted frequencies from predicted latent proportions once final counts are known \parencite{scoringprops}. Together, these advances enable robust real-time evaluation of model-based variant proportion predictions. We anticipate that providing these evaluation metrics and metadata will encourage further scientific investigation of nowcast outputs, ultimately improving the speed and robustness of model development.

Several limitations warrant discussion. First, this initial assessment includes only two individual modeling teams submitting probabilistic nowcasts for all locations, limiting the generalizability of performance conclusions from the probabilistic nowcast results. As additional teams have begun submitting following this assessment period, subsequent evaluations will include a greater number of models and enable more in-depth investigation of performance patterns. Second, the Hub's geographic scope is currently restricted to the US, despite global circulation of SARS-CoV-2 variants and documented patterns in country-to-country variant dynamics \parencite{Susswein2023, Molan2025}. While this decision aligns with the geographic scope of established forecasting Hubs \parencite{Mathis2024FluSight, FluSightHub, Reich2019Influenza, wolffram2023collaborative, cramer2022evaluation}, it may have biased initial participation towards U.S.-based modelers. Third, evaluation relies solely on publicly available GenBank data \parencite{benson2012genbank}, which represents only a partial view of U.S. sequencing efforts. While this constraint ensures equitable data access consistent with Hub conventions, it necessarily limits both the data easily available for model training and the scope of evaluation. Fourth, we used the scaled relative skill score computed across different strata to compare between different models with overlapping but not identical sets of nowcasts. However, it is not a proper scoring rule \parencite{Gneiting2007}, which by definition means that the metric, in this case the scaled relative skill score, does not always favor the true predictive distribution over all other possible distributions. The implications of this are that while scaled relative skill is a useful descriptive tool for comparing model performance, the metric may not consistently identify the best-calibrated model, and forecasters are not strictly incentivised to submit their true predictive distributions. We therefore treat scaled relative skill as a practical comparative measure rather than a theoretically grounded evaluation criterion, and supplement it with proper scoring rules, such as the energy score and Brier score, for model evaluation. Additionally, our analysis of the energy scores implicitly weights scores proportional to the number of reported sequences during the evaluation period. However, sequencing levels do not necessarily reflect what we care about most, but rather, where and when we have the most data to discern performance, with dates and locations with very few sequences having very little discerning observations regarding the true clade proportion. Finally, the Hub currently solicits estimates only for SARS-CoV-2 clades, though estimates of clade dynamics in other pathogens such as influenza and norovirus are also of public health relevance for understanding variant-driven surges \parencite{hay2025evaluation, Barclay2025} and improving vaccine strain selection \parencite{Huddleston2020Forecasting, Lee2025Reproducible, Krammer2023COVID}.

The collaborative Hub framework provides infrastructure for coordinated evaluation of pathogen evolution models. The infrastructure created here has the potential to speed up performance improvements of real-time variant nowcasting and forecasting models. Improvements in predictive accuracy, localized information, and longer-term projections of variant nowcast dynamics has the potential to support more localized policy and treatment recommendations, as well as potentially improving forecasting during variant-drive surges. In the future, variant nowcasting could provide warning signals of alarming variants which could drive surges in transmission and warrant further investigation, such as through lab-based studies to identify antigenic changes \parencite{planas2022considerable} or the creation of strain-specific PCR targets \parencite{Paulos2025, Buttinger2025} for variant monitoring. Nowcasts of variant proportions may be most valuable when observed data are sparse or noisy, as models can provide actionable estimates of variant dynamics that observed proportions alone cannot deliver. The evaluation methods and operational capacity described here provide pathways for expansion to other pathogens where variant dynamics drive disease burden, and we envision that this work could pave the way for future expansion to other countries and pathogens and providing a template for doing so rapidly in the case of an emerging outbreak.

In conclusion, this work describes the data provided by the U.S. SARS-CoV-2 Variant Nowcast Hub, the outputs solicited from submitting teams, the evaluation process, and an assessment of model performance over the Hub's first operational period. The Hub establishes reproducible infrastructure for real-time evaluation of pathogen evolution models. As participation expands and the sequencing landscapes evolve, we believe that the infrastructure described here will be useful for future outbreak response, enabling more timely and actionable information from genomic sequencing data. 

\section{Data availability}
All data supporting the findings of this study is available at the United States SARS-CoV-2 Variant Nowcast Hub at \url{https://github.com/reichlab/variant-nowcast-hub} and at Nextstrain \url{https://data.nextstrain.org/files/workflows/forecasts-ncov/open/nextstrain_clades/usa.tsv.gz}. 

\section{Code availability}
All code to reproduce results and figures is available publicly at \url{https://github.com/epiforecasts/evalvariantnowcasthub/}. A version DOI is available at \url{https://doi.org/10.5281/zenodo.20529200}. 

\section{Acknowledgments}
We acknowledge the financial support from CDC Grant NU38FT00008 (K.E.J., N.G.R., I.M., T.R., M.G., S.F., E.L.R., B.R., B.S.) and from NIGMS R35GM119582 (N.G.R.). This project was made possible by cooperative agreement NU38FT00008 from the CDC's Center for Forecasting and Outbreak Analytics. Its contents are solely the responsibility of the authors and do not necessarily represent the official views of the Centers for Disease Control and Prevention. We acknowledge the financial support from Wellcome 210758/Z/18/Z (S.F.). We acknowledge the financial support from the National Institutes of Health National Institute for Allergies and Infectious Diseases 75N93021C00015 (J.H. and J.L.). N.G.R. discloses a consulting relationship with Google LLC, although this funding and research are unrelated to the current publication. T.B. is a Howard Hughes Medical Institute Investigator. We thank Juan Castro, Scott Olesen, Clinton Paden from the U.S. Centers for Disease Control and Prevention for their input on the project.

\section{Author Contributions}
K.E.J., N.G.R., D.H.M., A.M., and E.L.R. conceived of the initial idea for the Variant Nowcast Hub. I.M., T.R. and K.E.J initially drafted the manuscript. I.M. and T.R. performed all programming relating to scoring model outputs and K.E.J performed all programming analyzing the data and outputs. N.G.R, M.G., E.L.R., B.R., B.S., I.M., and T.R. operationalized the Hub with the input from D.H.M, K.E.J and A.M. B.C., M.G., N.G.R. and N.L. contributed to the manuscript's writing. All other authors either submitted models to the Hub or provided input on the analyses presented in the manuscript. All authors gave final approval for submission. 

\section{Competing Interests statement}
N.G.R. reports consulting income from Google Research.

\section{References}
\printbibliography

@article{Hadfield2018,
  author = {Hadfield, James and Megill, Colin and Bell, Sidney M and Huddleston, John and Potter, Barney and Callender, Charlton and Sagulenko, Pavel and Bedford, Trevor and Neher, Richard A},
  title = {Nextstrain: real-time tracking of pathogen evolution},
  journal = {Bioinformatics},
  year = {2018},
  volume = {34},
  number = {23},
  pages = {4121--4123},
  doi = {10.1093/bioinformatics/bty407},
  url = {https://academic.oup.com/bioinformatics/article/34/23/4121/5001388}
}

@article{andrews2026nextstrain,
  title        = {Nextstrain automates real-time phylodynamic analysis of open data for endemic and emerging pathogens},
  author       = {Andrews, Kirsty R and Chang, James and Roemer, Cornelius and Hadfield, James and Lin, Victor and Brito, Anderson F and Daodu, Richard O and Joia, Ismail A and Kistler, Kelsey and Li, Albert and Moncla, Louise H and Paredes, Miguel I and K{\"u}hnert, Denise and Torres, Laura M and Voitl, Lukas and Aksamentov, Ivan and Hodcroft, Emma B and Huddleston, John and McCrone, John T and Anderson, Joel S J and Sibley, Thomas R and Lee, Justin and Neher, Richard A and Bedford, Trevor},
  year         = {2026},
  journal      = {bioRxiv},
  volume       = {2026.03.23.713807},
  doi          = {10.64898/2026.03.23.713807},
  note         = {Preprint}
}

@article{Shu2017,
  author = {Shu, Yuelong and McCauley, John},
  title = {{GISAID}: Global initiative on sharing all influenza data -- from vision to reality},
  journal = {Eurosurveillance},
  year = {2017},
  volume = {22},
  number = {13},
  article = {pii=30494},
  doi = {10.2807/1560-7917.ES.2017.22.13.30494}
}

@article{benson2012genbank,
  title={GenBank},
  author={Benson, Dennis A and Cavanaugh, Mark and Clark, Karen and Karsch-Mizrachi, Ilene and Lipman, David J and Ostell, James and Sayers, Eric W},
  journal={Nucleic acids research},
  volume={41},
  number={D1},
  pages={D36--D42},
  year={2012},
  publisher={Oxford University Press}
}

@article{Volz2021,
  author = {Volz, Erik and Mishra, Swapnil and Chand, Meera and Barrett, Jeffrey C. and Johnson, Robert and Geidelberg, Lily and Hinsley, Wes R. and Laydon, Daniel J. and Dabrera, Gavin and O'Toole, {\'A}ine and Amato, Robert and Ragonnet-Cronin, Manon and Harrison, Ian and Jackson, Ben and Ariani, Cristina V. and Boyd, Olivia and Loman, Nicholas J. and McCrone, John T. and Gon{\c{c}}alves, S{\'o}nia and Jorgensen, David and Myers, Richard and Hill, Verity and Jackson, David K. and Gaythorpe, Katy and Groves, Natalie and Sillitoe, John and Kwiatkowski, Dominic P. and {The COVID-19 Genomics UK (COG-UK) consortium} and Flaxman, Seth and Ratmann, Oliver and Bhatt, Samir and Hopkins, Susan and Gandy, Axel and Rambaut, Andrew and Ferguson, Neil M.},
  title = {Assessing transmissibility of {SARS-CoV-2} lineage {B.1.1.7} in {England}},
  journal = {Nature},
  year = {2021},
  volume = {593},
  pages = {266--269},
  doi = {10.1038/s41586-021-03470-x},
  url = {https://www.nature.com/articles/s41586-021-03470-x}
}

@article{abousamra2024fitness,
  title={Fitness models provide accurate short-term forecasts of {SARS-CoV-2} variant frequency},
  author={Abousamra, Eslam and Figgins, Marlin and Bedford, Trevor},
  journal={PLOS Computational Biology},
  volume={20},
  number={9},
  pages={e1012443},
  year={2024},
  publisher={Public Library of Science San Francisco, CA USA},
  doi={10.1371/journal.pcbi.1012443}
}

@misc{Variant-nowcast2024,
author = {{Reich Lab at UMass-Amherst}},
title  = {{US} {SARS}-{C}o{V}-2 {V}ariant {N}owcast {H}ub},
howpublished = {https://github.com/reichlab/variant-nowcast-hub},
year = {2024},
note = {Accessed: 2024-12-27}
}

@misc{hubverse2025github,
  author = {The Consortium of Infectious Disease Modeling Hubs},
  title  = {The hubverse: open tools for collaborative modeling},
  year  = {2025},
  howpublished = {\url{https://github.com/hubverse-org}},
  note         = {GitHub release v5.0.0, 17 Jan 2025, Accessed: 2025-06-13}
}

@article{rambaut2020dynamic,
  author  = {Rambaut, Andrew and Holmes, Edward C. and O'Toole, {\'A}ine and Hill, Verity and McCrone, John T. and Ruis, Carla and du Plessis, Louis and Pybus, Oliver G.},
  title   = {A dynamic nomenclature proposal for {SARS-CoV-2} lineages to assist genomic epidemiology},
  journal = {Nature Microbiology},
  year    = {2020},
  doi     = {10.1038/s41564-020-0770-5}
}

@misc{MMWR,
    author = {{Centers for Disease Control and Prevention }},
    title = {{MMWR Week Fact Sheet}},
    howpublished = {https://ndc.services.cdc.gov/wp-content/uploads/MMWR\_Week\_overview.pdf}
}

@misc{Roemer2024Tree,
author = { Roemer, Cornelius and Neher, Richard },
title = {{SARS-CoV-2} phylogeny},
year = {2024},
howpublished = {https://next.nextstrain.org/nextclade/sars-cov-2}
}

@article{Susswein2023,
	author = {Susswein, Zachary and Johnson, Kaitlyn E. and Kassa, Robel and Parastaran, Mina and Peng, Vivian and Wolansky, Leo and Scarpino, Samuel V. and Bento, Ana I.},
	title = {Leveraging global genomic sequencing data to estimate local variant dynamics},
	elocation-id = {2023.01.02.23284123},
	year = {2023},
	doi = {10.1101/2023.01.02.23284123},
	publisher = {Cold Spring Harbor Laboratory Press},
	URL = {https://www.medrxiv.org/content/early/2023/03/20/2023.01.02.23284123},
	eprint = {https://www.medrxiv.org/content/early/2023/03/20/2023.01.02.23284123.full.pdf},
	journal = {medRxiv}
}

@article{gneiting2008assessing,
  title={Assessing probabilistic forecasts of multivariate quantities, with an application to ensemble predictions of surface winds},
  author={Gneiting, Tilmann and Stanberry, Larissa I and Grimit, Eric P and Held, Leonhard and Johnson, Nicholas A},
  journal={Test},
  volume={17},
  pages={211--235},
  year={2008},
  publisher={Springer}
}

@article{jordan2019evaluating,
  title={Evaluating probabilistic forecasts with scoringRules},
  author={Jordan, Alexander and Kr{\"u}ger, Fabian and Lerch, Sebastian},
  journal={Journal of Statistical Software},
  volume={90},
  pages={1--37},
  year={2019}
}

@article{lopez2024challenges,
  title={Challenges of {COVID-19} Case Forecasting in the {US}, 2020--2021},
  author={Lopez, Velma K and Cramer, Estee Y and Pagano, Robert and Drake, John M and O'Dea, Eamon B and Adee, Madeline and Ayer, Turgay and Chhatwal, Jagpreet and Dalgic, Ozden O and Ladd, Mary A and others},
  journal={PLOS Computational Biology},
  volume={20},
  number={5},
  pages={e1011200},
  year={2024},
  publisher={Public Library of Science San Francisco, CA USA},
  doi={10.1371/journal.pcbi.1011200}
}

@article{daviesesttransmissibility2021,
author = {Nicholas G. Davies  and Sam Abbott  and Rosanna C. Barnard  and Christopher I. Jarvis  and Adam J. Kucharski  and James D. Munday  and Carl A. B. Pearson  and Timothy W. Russell  and Damien C. Tully  and Alex D. Washburne  and Tom Wenseleers  and Amy Gimma  and William Waites  and Kerry L. M. Wong  and Kevin van Zandvoort  and Justin D. Silverman  and {CMMID COVID-19 Working Group1\textdaggerdbl} and {COVID-19 Genomics UK (COG-UK) Consortium\textdaggerdbl} and Karla Diaz-Ordaz  and Ruth Keogh  and Rosalind M. Eggo  and Sebastian Funk  and Mark Jit  and Katherine E. Atkins  and W. John Edmunds },
title = {Estimated transmissibility and impact of SARS-CoV-2 lineage B.1.1.7 in England},
journal = {Science},
volume = {372},
number = {6538},
pages = {eabg3055},
year = {2021},
doi = {10.1126/science.abg3055},
URL = {https://www.science.org/doi/abs/10.1126/science.abg3055},
eprint = {https://www.science.org/doi/pdf/10.1126/science.abg3055}}

@article{Figgins2025, 
title={Inferring variant-specific effective reproduction numbers from combined case and sequencing data}, 
url={http://dx.doi.org/10.7554/eLife.104802.1}, 
DOI={10.7554/elife.104802.1}, 
journal={eLife Sciences Publications, Ltd}, 
author={Figgins, Marlin D. and Bedford, Trevor}, 
year={2025},
month=sep }

@article{Taylor2025, 
title={Founder effects arising from gathering dynamics systematically bias emerging pathogen surveillance}, 
url={http://dx.doi.org/10.7554/eLife.104201.1}, 
DOI={10.7554/elife.104201.1}, 
journal={eLife Sciences Publications, Ltd}, 
author={Taylor, Bradford P and Hanage, William P}, 
year={2025},
month=mar }

@misc{Figgins2024,
  author = {Figgins, Marlin D. and Bedford, Trevor},
  title = {Frequency dynamics predict viral fitness, antigenic relationships and epidemic growth},
  howpublished = {medRxiv preprint},
  year = {2024},
  doi = {10.1101/2024.12.02.24318334},
  url = {https://doi.org/10.1101/2024.12.02.24318334}
}

@software{pathoplexus2024,
  author = {{Pathoplexus}},
  title = {Pathoplexus},
  year = {2024},
  url = {https://pathoplexus.org},
  note = {Open-source database for human viral pathogen genomic data}
}

@article{Wadford2023,
  author = {Wadford, Debra A. and Baumrind, Nikki and Baylis, Elizabeth F. and Bell, John M. and Bouchard, Ellen L. and Crumpler, Megan and Foote, Eric M. and Gilliam, Sabrina and Glaser, Carol A. and Hacker, Jill K. and Ledin, Katya and Messenger, Sharon L. and Morales, Christina and Smith, Emily A. and Sevinsky, Joel R. and Corbett-Detig, Russell B. and DeRisi, Joseph and Jacobson, Kathleen and {the COVIDNet Consortium}},
  title = {Implementation of California COVIDNet -- a multi-sector collaboration for statewide SARS-CoV-2 genomic surveillance},
  journal = {Frontiers in Public Health},
  volume = {11},
  pages = {1249614},
  year = {2023},
  doi = {10.3389/fpubh.2023.1249614}
}

@article{Ray2023,
  author = {Ray, Evan L. and Brooks, Logan C. and Bien, Jacob and Biggerstaff, Matthew and Bosse, Nikos I. and Bracher, Johannes and Cramer, Estee Y. and Funk, Sebastian and Gerding, Aaron and Johansson, Michael A. and Rumack, Aaron and Wang, Yijin and Zorn, Martha and Tibshirani, Ryan J. and Reich, Nicholas G.},
  title = {Comparing trained and untrained probabilistic ensemble forecasts of COVID-19 cases and deaths in the United States},
  journal = {International Journal of Forecasting},
  year = {2023},
  volume = {39},
  pages = {1366--1383},
  doi = {10.1016/j.ijforecast.2022.06.005}
}

@misc{nextstrainforecasts,
  author = {Hadfield, James and Megill, Colin and Bell, Sidney M. and Huddleston, John and Potter, Barney and Callender, Charlton and Sagulenko, Pavel and Bedford, Trevor and Neher, Richard A.},
  title = {Nextstrain: SARS-CoV-2 Forecasts},
  howpublished = {\url{https://nextstrain.org/sars-cov-2/forecasts}},
  note = {Accessed: 2026-01-16},
  year = {2024}
}

@misc{scoringprops,
author = {Robacker, Thomas and MacArthur, Isaac and Morris, Dylan and Johnson, Kaitlyn E. and Reich, Nicholas G. and Griffin, Maryclare},
title = {Evaluating predictions of multi-class population proportions: A variant nowcasting case study.},
howpublished = {work in progress}
}

@Manual{scoringutils,
  title = {scoringutils: Utilities for Scoring and Assessing Predictions},
  author = {Nikos I. Bosse and Hugo Gruson and Sebastian Funk and EpiForecasts and Sam Abbott},
  year = {2020},
  doi = {10.5281/zenodo.4618017},
  url = {https://cran.r-project.org/package=scoringutils}
}

@article{Reichmuth2023,
  author = {Reichmuth, Martina L. and Hodcroft, Emma B. and Althaus, Christian L.},
  title = {Importation of Alpha and Delta variants during the SARS-CoV-2 epidemic in Switzerland: Phylogenetic analysis and intervention scenarios},
  journal = {PLoS Pathogens},
  volume = {19},
  number = {8},
  pages = {e1011553},
  year = {2023},
  month = {8},
  doi = {10.1371/journal.ppat.1011553}
}

@article{hubverse2025,
  author = {{Consortium of Infectious Disease Modeling Hubs} and Kerr, Melissa and Borchering, Rebecca and Castro Rivadeneira, Alvaro and Contamin, Lucie and Funk, Sebastian and Hochheiser, Harry and Howerton, Emily and Krystalli, Anna and Shandross, Li and Reich, Nicholas G},
  title = {Coordinating collaborative infectious disease modeling projects with the hubverse},
  journal = {medRxiv},
  year = {2025},
  doi = {10.1101/2025.10.03.25337284},
  url = {https://www.medrxiv.org/content/10.1101/2025.10.03.25337284v1},
  note = {Preprint}
}

@article{Feng2024,
  author = {Feng, Yinan and Goldberg, Emma E. and Kupperman, Michael and Zhang, Xitong and Lin, Youzuo and Ke, Ruian},
  title = {{CovTransformer}: A transformer model for {SARS-CoV-2} lineage frequency forecasting},
  journal = {Virus Evolution},
  volume = {10},
  number = {1},
  pages = {veae086},
  year = {2024},
  doi = {10.1093/ve/veae086},
  url = {https://doi.org/10.1093/ve/veae086}
}

@article{Bracher2021,
  author = {Bracher, Johannes and Ray, Evan L. and Gneiting, Tilmann and Reich, Nicholas G.},
  title = {Evaluating epidemic forecasts in an interval format},
  journal = {PLOS Computational Biology},
  year = {2021},
  volume = {17},
  number = {2},
  pages = {e1008618},
  doi = {10.1371/journal.pcbi.1008618}
}

@article{Reich2019Influenza,
  author = {Reich, Nicholas G. and Brooks, Logan C. and Fox, Spencer J. and Kandula, Sasikiran and McGowan, Craig J. and Moore, Evan and Osthus, Dave and Ray, Evan L. and Tushar, Abhinav and Yamana, Teresa K. and Biggerstaff, Matthew and Johansson, Michael A. and Rosenfeld, Roni and Shaman, Jeffrey},
  title = {A collaborative multiyear, multimodel assessment of seasonal influenza forecasting in the United States},
  journal = {Proceedings of the National Academy of Sciences},
  year = {2019},
  volume = {116},
  number = {8},
  pages = {3146--3154},
  doi = {10.1073/pnas.1812594116}
}

@article{Chen2022,
  author = {Chen, Zugen and Azman, Andrew S. and Chen, Xi and Zou, Jin and Tian, Yu and Sun, Ruijun and Xu, Xiang and Wu, Yuyang and Lu, Wanying and Ge, Sheng and Zhao, Zheng and Yang, Jing and Leung, Daniel T. and Luquero, Francisco J. and Tian, Dayan and Hung, Curtis T. and Lu, Yun and Jiang, Xin},
  title = {Global landscape of SARS-CoV-2 genomic surveillance and data sharing},
  journal = {Nature Genetics},
  year = {2022},
  volume = {54},
  pages = {499--507},
  doi = {10.1038/s41588-022-01033-y},
  note = {Published online 28 March 2022}
}

@article{Park2023,
author = {Sang Woo Park  and Kaiyuan Sun  and Sam Abbott  and Ron Sender  and Yinon M. Bar-on  and Joshua S. Weitz  and Sebastian Funk  and Bryan T. Grenfell  and Jantien A. Backer  and Jacco Wallinga  and Cecile Viboud  and Jonathan Dushoff },
title = {Inferring the differences in incubation-period and generation-interval distributions of the Delta and Omicron variants of SARS-CoV-2},
journal = {Proceedings of the National Academy of Sciences},
volume = {120},
number = {22},
pages = {e2221887120},
year = {2023},
doi = {10.1073/pnas.2221887120},
URL = {https://www.pnas.org/doi/abs/10.1073/pnas.2221887120},
eprint = {https://www.pnas.org/doi/pdf/10.1073/pnas.2221887120}
}

@article{tegally2022emergence,
  title={Emergence of SARS-CoV-2 Omicron lineages BA.4 and BA.5 in South Africa},
  author={Tegally, Houriiyah and Moir, Monika and Everatt, Josie and Giovanetti, Marta and Scheepers, Cathrine and Wilkinson, Eduan and Subramoney, Kathleen and Makatini, Zinhle and Moyo, Sikhulile and Amoako, Daniel G and Baxter, Chad and Althaus, Christian L and Anyaneji, Ugochukwu J and Kekana, Dikeledi and Viana, Raquel and Giandhari, Jennifer and Maponga, Thando G and Maruapula, Dorcas and Choga, Wonderful and Matshaba, Mogomotsi and Mbhele, Nokuzola and Msomi, Nokukhanya and Naidoo, Yeshnee and Pillay, Sureshnee and Sanko, Timothy J and San, James E and Scott, Linda and Singh, Lavanya and Magini, Noluthando A and Smith-Lawrence, Pamela and Stevens, Wendy and Dor, Gert and Tshiabuila, Derek and Wolter, Nicole and Preiser, Wolfgang and Treurnicht, Florette K and Venter, Marietjie and Chiloane, Gert and McIntyre, Cheryl and O'Toole, {\'A}ine and Rambaut, Andrew and Martin, Darren P and Williamson, Carolyn and Hsiao, Nei-yuan and von Gottberg, Anne and Bhiman, Jinal N and Lessells, Richard J and de Oliveira, Tulio},
  journal={Nature Medicine},
  volume={28},
  number={9},
  pages={1785--1790},
  year={2022},
  month={9},
  publisher={Nature Publishing Group},
  doi={10.1038/s41591-022-01911-2},
  pmid={35732823},
  pmcid={PMC9470493}
}

@article{Paulos2025,
  author = {Paulos, Abigail P. and Hilton, Stephen P. and Shelden, Bridgette and Duong, Dorothea and Boehm, Alexandria B. and Wolfe, Marlene K.},
  title = {Detection of Hemagglutinin {H5} Influenza {A} Virus {RNA} and Model of Potential Inputs in an Urban {California} Sewershed},
  journal = {Environmental Science \& Technology},
  volume = {59},
  number = {31},
  pages = {16168--16179},
  year = {2025},
  doi = {10.1021/acs.est.4c14792}
}

@article{Buttinger2025,
  author = {Buttinger, Gerhard and Petrillo, Mauro and Valastro, Viviana and Marciano, Sabrina and Crimaudo, Marika and D'Amico, Valeria and Leoni, Gabriele and Seigneuric, Renaud and Paracchini, Valentina and Robouch, Piotr and Lambrecht, B{\'e}n{\'e}dicte and Gawlik, Bernd Manfred and Terregino, Calogero and Veneri, Carolina and La Rosa, Giuseppina and Suffredini, Elisabetta and Querci, Maddalena and Panzarin, Valentina and Marchini, Antonio},
  title = {Novel (d)PCR assays for influenza A(H5Nx) viruses clade 2.3.4.4b surveillance},
  journal = {Eurosurveillance},
  volume = {30},
  number = {33},
  pages = {pii=2500183},
  year = {2025},
  doi = {10.2807/1560-7917.ES.2025.30.33.2500183},
  url = {https://doi.org/10.2807/1560-7917.ES.2025.30.33.2500183},
  note = {Received: 12 Mar 2025; Accepted: 26 Jun 2025}
}

@article{Gneiting2007,
  author  = {Gneiting, Tilmann and Raftery, Adrian E.},
  title   = {Strictly Proper Scoring Rules, Prediction, and Estimation},
  journal = {Journal of the American Statistical Association},
  year    = {2007},
  volume  = {102},
  number  = {477},
  pages   = {359--378},
  doi     = {10.1198/016214506000001437}
}

@article{planas2022considerable,
  title={Considerable escape of SARS-CoV-2 Omicron to antibody neutralization},
  author={Planas, Delphine and Saunders, Nell and Maes, Piet and Guivel-Benhassine, Florence and Planchais, Cyril and Buchrieser, Julian and Bolland, Wai-Hung and Porrot, Fran{\c{c}}oise and Staropoli, Isabelle and Lemoine, Fran{\c{c}}ois and others},
  journal={Nature},
  volume={602},
  number={7898},
  pages={671--675},
  year={2022},
  publisher={Nature Publishing Group UK London}
}

@misc{cladetime,
  author = {Sweger, Becky and Ray, Evan and Rogers, ben and Kamvar, Zhian and Cornell, Mattew and Reich, Nicholas G.},
  title = {cladetime: Python interface for accessing {Nextstrain} {SARS-CoV-2} sequence and clade data},
  year = {2024},
  howpublished = {Python Package Index},
  url = {https://github.com/reichlab/cladetime},
  note = {Available at https://pypi.org/project/cladetime/}
}

@misc{hay2025evaluation,
  title={Evaluation of the epidemiological outlook of the influenza A/H3N2 clade K in England during the 2025-26 season},
  author={Hay, James A and Alahakoon, Punya and Greenshields-Watson, Alexander and Kendall, Michelle and Ghafari, Mahan and Wymant, Chris and Hinch, Robert and Ferretti, Luca and Panovska-Griffiths, Jasmina and Fraser, Christophe},
  year={2025},
  month={11},
  publisher={Zenodo},
  doi={10.5281/zenodo.17704679},
  howpublished={Preprint. \url{https://doi.org/10.5281/zenodo.17704679}}
}

@misc{FluSightHub,
  author = {{CDC FluSight Team}},
  title = {{FluSight} Forecast Hub: Repository for Forecasts of Influenza Hospitalizations},
  year = {2023},
  howpublished = {GitHub repository},
  url = {https://github.com/cdcepi/FluSight-forecast-hub},
  note = {Accessed: 2026-01-12}
}

@article{Mathis2024FluSight,
  author = {Mathis, Sarabeth M. and Webber, Alexander E. and Le{\'o}n, Tiffany M. and Murray, Emily L. and Sun, Madeline and White, Lindsay A. and Brooks, Logan C. and Green, Alden and Hu, A. J. and Koppaka, Rangarao and Segaloff, Hailey E. and Slayton, Rachel B. and Johansson, Michael A. and Reich, Nicholas G. and Biggerstaff, Matthew and Borchering, Rebecca K.},
  title = {Evaluation of FluSight influenza forecasting in the 2021--22 and 2022--23 seasons with a new target laboratory-confirmed influenza hospitalizations},
  journal = {Nature Communications},
  year = {2024},
  volume = {15},
  pages = {6289},
  doi = {10.1038/s41467-024-50601-9}
}

@article{Ragonnet-Cronin2023,
  author = {Ragonnet-Cronin, Manon and Nutalai, Rungtiwa and Huo, Jiandong and Dijokaite-Guraliuc, Aiste and Das, Raksha and Tuekprakhon, Aekkachai and Supasa, Piyada and Liu, Chang and Selvaraj, Muneeswaran and Groves, Natalie and Hartman, Hassan and Ellaby, Nicholas and Sutton, J. Mark and Bahar, Mohammad W. and Zhou, Daming and Fry, Elizabeth and Ren, Jingshan and Brown, Colin and Klenerman, Paul and Dunachie, Susanna J. and Mongkolsapaya, Juthathip and Hopkins, Susan and Chand, Meera and Stuart, David I. and Screaton, Gavin R. and Rokadiya, Sakib},
  title = {Generation of SARS-CoV-2 escape mutations by monoclonal antibody therapy},
  journal = {Nature Communications},
  year = {2023},
  volume = {14},
  number = {1},
  pages = {3334},
  doi = {10.1038/s41467-023-37826-w},
  url = {https://doi.org/10.1038/s41467-023-37826-w}
}

@misc{CDCVariantNowcast,
  author = {{Centers for Disease Control and Prevention}},
  title = {SARS-CoV-2 Variants and Genomic Surveillance},
  howpublished = {\url{https://www.cdc.gov/covid/php/variants/variants-and-genomic-surveillance.html}},
  note = {Accessed: 2026-01-11},
  year = {2024}
}

@misc{CoVSpectrum,
  author = {Chen, Chaoran and Nadeau, Sarah and Yared, Michael and Voinov, Philipp and Xie, Ning and Roemer, Cornelius and Stadler, Tanja},
  title = {CoV-Spectrum: Analysis of Globally Shared SARS-CoV-2 Data to Identify and Characterize New Variants},
  howpublished = {\url{https://cov-spectrum.org/explore/United\%20Kingdom/AllSamples/Past6M}},
  note = {Accessed: 2026-01-11},
  year = {2022}
}

@misc{UKHSATechnicalBriefing38,
  author = {{UK Health Security Agency}},
  title = {SARS-CoV-2 variants of concern and variants under investigation in England: Technical briefing 38},
  institution = {UK Health Security Agency},
  year = {2022},
  month = {3},
  type = {Technical Briefing},
  number = {38},
  url = {https://assets.publishing.service.gov.uk/media/622b4a20e90e070ed8233a05/Technical-Briefing-38-11March2022.pdf},
  note = {Published: 11 March 2022}
}

@misc{COGUKConsortium,
  author = {{COVID-19 Genomics UK (COG-UK) Consortium}},
  title = {COG-UK Consortium},
  howpublished = {\url{https://www.sanger.ac.uk/collaboration/covid-19-genomics-uk-cog-uk-consortium/}},
  note = {Accessed: 2026-01-11},
  year = {2020},
  organization = {Wellcome Sanger Institute}
}

@article{Paul2021,
  author = {Paul, Prabasaj and France, Anne Marie and Aoki, Yutaka and Batra, Dhwani and Biggerstaff, Matthew and Dugan, Vivien and Galloway, Summer and Hall, Aron J. and Johansson, Michael A. and Kondor, Rebecca J. and Laufer Halpin, Alison and Lee, Brian and Lee, Justin S. and Limbago, Brandi and MacNeil, Adam and MacCannell, Duncan and Paden, Clinton R. and Queen, Krista and Reese, Heather E. and Retchless, Adam C. and Slayton, Rachel B. and Steele, Molly and Tong, Suxiang and Walters, Maroya S. and Wentworth, David E. and Silk, Benjamin J.},
  title = {Genomic Surveillance for SARS-CoV-2 Variants Circulating in the United States, December 2020--May 2021},
  journal = {MMWR. Morbidity and Mortality Weekly Report},
  year = {2021},
  volume = {70},
  number = {23},
  pages = {846--850},
  doi = {10.15585/mmwr.mm7023a3},
  pmid = {34111060},
  pmcid = {PMC8191868}
}

@article{Huddleston2020Forecasting,
  author = {Huddleston, John and Barnes, John R. and Rowe, Thomas and Xu, Xiyan and Kondor, Rebecca and Wentworth, David E. and Whittaker, Lynne and Ermetal, Burcu and Daniels, Rodney S. and McCauley, John W. and Fujisaki, Seiichiro and Nakamura, Kazuya and Kishida, Noriko and Watanabe, Shinji and Hasegawa, Hideki and Barr, Ian and Subbarao, Kanta and Barrat-Charlaix, Pierre and Neher, Richard A. and Bedford, Trevor},
  title = {Integrating genotypes and phenotypes improves long-term forecasts of seasonal influenza A/H3N2 evolution},
  journal = {eLife},
  year = {2020},
  volume = {9},
  pages = {e60067},
  doi = {10.7554/eLife.60067}
}

@article{Lee2025Reproducible,
  author = {Lee, Sojung and Viboud, C{\'e}cile and Bedford, Trevor},
  title = {Reproducible and later vaccine strain selection can improve vaccine match to A/H3N2 seasonal influenza viruses},
  journal = {npj Vaccines},
  year = {2025},
  volume = {10},
  pages = {23},
  doi = {10.1038/s41541-025-01292-w}
}

@article {Molan2025,
	author = {Molan, S. and Smith, N.K. and Gandhi, V. and Li, M. and Colijn, C. and Murall, C.L. and Stockdale, J.E.},
	title = {Drivers of COVID-19 variant wave dynamics: inferring oncoming wave size using global data with genomics},
	elocation-id = {2025.09.16.25335896},
	year = {2025},
	doi = {10.1101/2025.09.16.25335896},
	publisher = {Cold Spring Harbor Laboratory Press},
	URL = {https://www.medrxiv.org/content/early/2025/09/18/2025.09.16.25335896},
	eprint = {https://www.medrxiv.org/content/early/2025/09/18/2025.09.16.25335896.full.pdf},
	journal = {medRxiv}
}

@article{Campbell2019Bayesian,
  author = {Campbell, Finlay and Cori, Anne and Ferguson, Neil and Jombart, Thibaut},
  title = {Bayesian inference of transmission chains using timing of symptoms, pathogen genomes and contact data},
  journal = {PLOS Computational Biology},
  year = {2019},
  volume = {15},
  number = {3},
  pages = {e1006930},
  doi = {10.1371/journal.pcbi.1006930}
}

@article{Hill2022Phylogenetics,
  author = {Hill, Verity and Du Plessis, Louis and Peacock, Thomas P. and Aggarwal, Dinesh and Ashford, Fiona and Bibi, Sagida and Bridges, Emma and Brown, Julianne R. and Brunker, Kirstyn and Bull, Matthew and Carden, Holli and Dabrera, Gavin and Dabrera, Gavin and Dabrera, Gavin and Dabrera, Gavin and Dabrera, Gavin and Rambaut, Andrew and {COVID-19 Genomics UK (COG-UK) Consortium}},
  title = {The origins and molecular evolution of {SARS-CoV-2} lineage {B.1.1.7} in the {UK}},
  journal = {Virus Evolution},
  year = {2022},
  volume = {8},
  number = {2},
  pages = {veac080},
  doi = {10.1093/ve/veac080}
}

@article{Giardina2022Methods,
  author = {Giardina, Fabr{\`i}cia and Romero-Severson, Ethan O. and Albert, Jan and Leitner, Thomas},
  title = {Methods combining genomic and epidemiological data in the reconstruction of transmission trees: a systematic review},
  journal = {Pathogens},
  year = {2022},
  volume = {11},
  number = {2},
  pages = {252},
  doi = {10.3390/pathogens11020252}
}

@article{Crits-Christoph2024Market,
  author = {Crits-Christoph, Alexander and Levy, Joshua I. and Pekar, Jonathan E. and Goldstein, Stephen A. and Singh, Reema and Hensel, Zach and Gangavarapu, Karthik and Moshiri, Niema and Havens, Jennifer L. and Gangavarapu, Karthik and Suchard, Marc A. and Koopmans, Marion P. G. and Lemey, Philippe and Wertheim, Joel O. and Worobey, Michael and Holmes, Edward C. and Andersen, Kristian G. and D{\'e}barre, Florence},
  title = {Genetic tracing of market wildlife and viruses at the epicenter of the {COVID-19} pandemic},
  journal = {Cell},
  year = {2024},
  volume = {187},
  number = {19},
  pages = {5468--5482.e11},
  doi = {10.1016/j.cell.2024.08.010}
}

@article{Krammer2023COVID,
  author = {Krammer, Florian and Rouphael, Nadine},
  title = {{SARS-CoV-2} vaccine strain selection: {Guidance} from influenza},
  journal = {Open Forum Infectious Diseases},
  year = {2023},
  volume = {10},
  number = {6},
  pages = {ofad239},
  doi = {10.1093/ofid/ofad239}
}

@article{Barclay2025,
  author = {Barclay, Leslie and Montmayeur, Anna M and Cannon, Jennifer L and Mallory, Michael L and Reyes, Yaoska I and Wall, Helen and Baric, Ralph S and Lindesmith, Lisa C and Vinj{\'e}, Jan and Chhabra, Preeti},
  title = {Molecular Evolution and Epidemiology of Norovirus GII.4 Viruses in the United States},
  journal = {The Journal of Infectious Diseases},
  volume = {232},
  number = {4},
  pages = {933--942},
  year = {2025},
  month = {10},
  issn = {1537-6613},
  doi = {10.1093/infdis/jiaf100},
  url = {https://doi.org/10.1093/infdis/jiaf100},
}

@article{Lorenz2025,
  author = {Lorenz, Camila and Lemey, Philippe and Dellicour, Simon},
  title = {Dengue serotypes and epidemic dynamics in Brazil: a spatiotemporal perspective},
  journal = {Travel Medicine and Infectious Disease},
  year = {2025},
  volume = {68},
  pages = {102948},
  doi = {10.1016/j.tmaid.2025.102948},
  url = {https://doi.org/10.1016/j.tmaid.2025.102948},
}

@article{Aksamentov2021,
  author = {Aksamentov, Ivan and Roemer, Cornelius and Hodcroft, Emma B. and Neher, Richard A.},
  title = {Nextclade: clade assignment, mutation calling and quality control for viral genomes},
  journal = {Journal of Open Source Software},
  volume = {6},
  number = {67},
  pages = {3773},
  year = {2021},
  publisher = {The Open Journal},
  doi = {10.21105/joss.03773},
  url = {https://doi.org/10.21105/joss.03773},
}

@article{Johnson2021,
  author = {Johnson, Kaitlyn E and Woody, Spencer and Lachmann, Michael and Fox, Spencer J and Klima, Jessica and Hines, Terrance S and Meyers, Lauren Ancel},
  title = {Real-Time Projections of SARS-CoV-2 B.1.1.7 Variant in a University Setting, Texas, USA},
  journal = {Emerging Infectious Diseases},
  volume = {27},
  number = {12},
  pages = {3188--3190},
  year = {2021},
  month = {12},
  doi = {10.3201/eid2712.210652},
  pmid = {34708684},
  pmcid = {PMC8632165},
  url = {https://doi.org/10.3201/eid2712.210652},
}

@article{McCrone2025,
  author = {McCrone, John T and Baele, Guy and Omah, Ifeanyi F and Kinganda-Lusamaki, Eddy and Brew, Joseph A and Carvalho, Luiz M and Dudas, Gytis and Mbala-Kingebeni, Placide and Suchard, Marc A and Rambaut, Andrew},
  title = {Evidence of latency reshapes our understanding of Ebola virus reservoir dynamics},
  journal = {bioRxiv},
  year = {2025},
  month = {10},
  doi = {10.1101/2025.10.17.683141},
  url = {https://www.biorxiv.org/content/10.1101/2025.10.17.683141v1},
  note = {Preprint},
}

@article{Srivastava2025,
  author = {Srivastava, Shriyansh and Sharma, Dheeraj and Sridhar, Sathvik Belagodu and Kumar, Sachin and Rao, G S N Koteswara and Budha, Roja Rani and Babu, Molakpogu Ravindra and Sahu, Rakesh and Sah, Sanjit and Mehta, Rachana and Giraldo-Corrales, Nahun Alejandro and Feehan, Jack and Apostolopoulos, Vasso and Rodriguez-Morales, Alfonso J},
  title = {Comparative analysis of Mpox clades: epidemiology, transmission dynamics, and detection strategies},
  journal = {BMC Infectious Diseases},
  volume = {25},
  pages = {1290},
  year = {2025},
  month = {10},
  doi = {10.1186/s12879-025-11784-8},
  pmid = {41083957},
  pmcid = {PMC12516921},
  url = {https://pmc.ncbi.nlm.nih.gov/articles/PMC12516921/},
}

@article{wolffram2023collaborative,
  title={Collaborative nowcasting of {COVID-19} hospitalization incidences in {Germany}},
  author={Wolffram, Daniel and Abbott, Sam and an der Heiden, Matthias and Funk, Sebastian and G{\"u}nther, Felix and Hailer, Davide and Heyder, Stefan and Hotz, Thomas and van de Kassteele, Jan and K{\"u}chenhoff, Helmut and M{\"u}ller-Hansen, S{\"o}ren and Syliqi, Diel{\"e} and Ullrich, Alexander and Weigert, Maximilian and Schienle, Melanie and Bracher, Johannes},
  journal={PLOS Computational Biology},
  volume={19},
  number={8},
  pages={e1011394},
  year={2023},
  publisher={Public Library of Science San Francisco, CA USA},
  doi={10.1371/journal.pcbi.1011394}
}

@article{cramer2022evaluation,
  title={Evaluation of individual and ensemble probabilistic forecasts of {COVID-19} mortality in the {United States}},
  author={Cramer, Estee Y and Ray, Evan L and Lopez, Velma K and Bracher, Johannes and Brennen, Andrea and Castro Rivadeneira, Alvaro J and Gerding, Aaron and Gneiting, Tilmann and House, Katie H and Huang, Yuxin and others},
  journal={Proceedings of the National Academy of Sciences},
  volume={119},
  number={15},
  pages={e2113561119},
  year={2022},
  publisher={National Academy of Sciences},
  doi={10.1073/pnas.2113561119}
}

@misc{reich2022collaborative,
  title={Collaborative hubs: making the most of predictive epidemic modeling},
  author={Reich, Nicholas G and Lessler, Justin and Funk, Sebastian and Viboud, Cecile and Vespignani, Alessandro and Tibshirani, Ryan J and Shea, Katriona and Schienle, Melanie and Runge, Michael C and Rosenfeld, Roni and others},
  journal={American Journal of Public Health},
  volume={112},
  number={6},
  pages={839--842},
  year={2022},
  publisher={American Public Health Association}
}

@article{Telenti2022,
  title={The evolution and biology of SARS-CoV-2 variants},
  author={Telenti, Amalio and Hodcroft, Emma B and Robertson, David L},
  journal={Cold Spring Harbor perspectives in medicine},
  volume={12},
  number={5},
  pages={a041390},
  year={2022},
  publisher={Cold Spring Harbor Laboratory Press}
}

\end{document}